\documentclass[twocolumn]{aastex63}

\usepackage{amsmath}
\usepackage{lineno}

\graphicspath{{./}{figures/}}

\renewcommand{\P}[0]{\mathcal{P}}

\shorttitle{COMAP Early Science: CO Data Processing}
\shortauthors{Foss et al.}

\begin{document}

\title{COMAP Early Science: III. CO Data Processing}

\correspondingauthor{Marie K. Foss}
\email{m.k.foss@astro.uio.no}

\author[0000-0001-8896-3159]{Marie K. Foss}
\author[0000-0003-3420-7766]{H\aa vard T. Ihle}
\author{Jowita Borowska}
\affil{Institute of Theoretical Astrophysics, University of Oslo, P.O. Box 1029 Blindern, N-0315 Oslo, Norway}

\author[0000-0002-8214-8265]{Kieran A. Cleary}
\affil{California Institute of Technology, Pasadena, CA 91125, USA}

\author[0000-0003-2332-5281]{Hans Kristian Eriksen}
\affil{Institute of Theoretical Astrophysics, University of Oslo, P.O. Box 1029 Blindern, N-0315 Oslo, Norway}

\author[0000-0001-7911-5553]{Stuart E.~Harper}
\affil{Jodrell Bank Centre for Astrophysics, Alan Turing Building, Department of Physics and Astronomy, School of Natural Sciences, The University of Manchester, Oxford Road, Manchester, M13 9PL, U.K.}
\author[0000-0002-4274-9373]{Junhan Kim}
\affil{California Institute of Technology, Pasadena, CA 91125, USA}
\author[0000-0002-5959-1285]{James W. Lamb}
\affil{Owens Valley Radio Observatory, California Institute of Technology, Big Pine, CA 93513, USA}
\author{Jonas G. S. Lunde}
\affil{Institute of Theoretical Astrophysics, University of Oslo, P.O. Box 1029 Blindern, N-0315 Oslo, Norway}

\author[0000-0001-7612-2379]{Liju Philip}
\affil{Jet Propulsion Laboratory, California Institute of Technology, 4800 Oak Grove Drive, Pasadena, CA 91109, USA}
\author{Maren Rasmussen}
\author[0000-0001-5301-1377]{Nils-Ole Stutzer}
\affil{Institute of Theoretical Astrophysics, University of Oslo, P.O. Box 1029 Blindern, N-0315 Oslo, Norway}
\author[0000-0001-8526-3464]{Bade D.~Uzgil}
\affiliation{California Institute of Technology, 1200 E. California Blvd., Pasadena, CA 91125, USA}
\author[0000-0002-5437-6121]{Duncan J. Watts}
\author[0000-0003-3821-7275]{Ingunn K. Wehus}
\affil{Institute of Theoretical Astrophysics, University of Oslo, P.O. Box 1029 Blindern, N-0315 Oslo, Norway}
\author{David P. Woody}
\affil{Owens Valley Radio Observatory, California Institute of Technology, Big Pine, CA 93513, USA}
\author[0000-0003-2358-9949 ]{J. Richard Bond}
\affil{Canadian Institute for Theoretical Astrophysics, University of Toronto, 60 St. George Street, Toronto, ON M5S 3H8, Canada}
\author[0000-0001-8382-5275]{Patrick C. Breysse}
\affil{Center for Cosmology and Particle Physics, Department of Physics, New York University, 726 Broadway, New York, NY, 10003, USA}
\author{Morgan Catha}
\affil{Owens Valley Radio Observatory, California Institute of Technology, Big Pine, CA 93513, USA}
\author{Sarah E.~Church}
\affiliation{Kavli Institute for Particle Astrophysics and Cosmology \& Physics Department, Stanford University, Stanford, CA 94305, USA}
\author[0000-0003-2618-6504]{Dongwoo T.~Chung}
\affil{Canadian Institute for Theoretical Astrophysics, University of Toronto, 60 St. George Street, Toronto, ON M5S 3H8, Canada}
\affiliation{Dunlap Institute for Astronomy and Astrophysics, University of Toronto, 50 St. George Street, Toronto, ON M5S 3H4, Canada}
\author[0000-0002-0045-442X]{Clive Dickinson}
\affil{Jodrell Bank Centre for Astrophysics, Alan Turing Building, Department of Physics and Astronomy, School of Natural Sciences, The University of Manchester, Oxford Road, Manchester, M13 9PL, U.K.}
\author[0000-0002-5223-8315]{Delaney A.~Dunne}
\affil{California Institute of Technology, Pasadena, CA 91125, USA}
\author{Todd Gaier}
\affiliation{Jet Propulsion Laboratory, California Institute of Technology, 4800 Oak Grove Drive, Pasadena, CA 91109, USA}
\author{Joshua Ott Gundersen}
\affil{Department of Physics, University of Miami, 1320 Campo Sano Avenue, Coral Gables, FL 33146, USA}
\author[0000-0001-6159-9174]{Andrew I.~Harris}
\affiliation{Department of Astronomy, University of Maryland, College Park, MD 20742, USA}
\author{Richard Hobbs}
\affiliation{Owens Valley Radio Observatory, California Institute of Technology, Big Pine, CA 93513, USA}
\author{Charles R.~Lawrence}
\affiliation{Jet Propulsion Laboratory, California Institute of Technology, 4800 Oak Grove Drive, Pasadena, CA 91109, USA}
\author{Norman Murray}
\affil{Canadian Institute for Theoretical Astrophysics, University of Toronto, 60 St. George Street, Toronto, ON M5S 3H8, Canada}
\author[0000-0001-9152-961X]{Anthony C.~S. Readhead}
\affil{California Institute of Technology, Pasadena, CA 91125, USA}
\author[0000-0002-8800-5740]{Hamsa Padmanabhan}
\affil{Departement de Physique Théorique, Universite de Genève, 24 Quai Ernest-Ansermet, CH-1211 Genève 4, Switzerland}
\author[0000-0001-5213-6231]{Timothy J.~Pearson}
\affil{California Institute of Technology, Pasadena, CA 91125, USA}
\author[0000-0002-1667-3897]{Thomas J. Rennie}
\affil{Jodrell Bank Centre for Astrophysics, Alan Turing Building, Department of Physics and Astronomy, School of Natural Sciences, The University of Manchester, Oxford Road, Manchester, M13 9PL, U.K.}
\collaboration{35}{(COMAP Collaboration)}

\begin{abstract}

  We describe the first season COMAP analysis pipeline that converts raw detector readouts to calibrated sky maps. This pipeline implements four main steps: gain calibration, filtering, data selection, and map-making. Absolute gain calibration relies on a combination of instrumental and astrophysical sources, while relative gain calibration exploits real-time total-power variations. High efficiency filtering is achieved through spectroscopic common-mode rejection within and across receivers, resulting in nearly uncorrelated white noise within single-frequency channels. Consequently, near-optimal but biased maps are produced by binning the filtered time stream into pixelized maps; the corresponding signal bias transfer function is estimated through simulations. Data selection is performed automatically through a series of goodness-of-fit statistics, including $\chi^2$ and multi-scale correlation tests. 
  Applying this pipeline to the first-season COMAP data, we produce a dataset with very low levels of correlated noise. We find that one of our two scanning strategies (the Lissajous type) is sensitive to residual instrumental systematics. As a result, we no longer use this type of scan and exclude data taken this way from our Season 1 power spectrum estimates. We perform a careful analysis of our data processing and observing efficiencies and take account of planned improvements to estimate our future performance. Power spectrum results derived from the first-season COMAP maps are presented and discussed in companion papers.
\end{abstract}


\section{Introduction} \label{sec:intro}
Understanding the evolution of galaxies and the intergalactic medium (IGM) over the largest spatial and temporal scales is one of the principal goals of cosmology. Galaxy surveys address this challenge by resolving and detecting individual galaxies, a technique that necessarily favors brighter galaxies and smaller cosmic volumes. Spectral line intensity mapping (LIM) \citep{madau1997, Battye2004, Peterson2006, Loeb2008} is a complementary technique (see \citealt{Kovetz+2017} or \citealt{Kovetz+2019} for a review) that holds the potential to characterize the global properties of galaxies and their evolution by surveying the aggregate emission from all galaxies over large volumes. 

This technique uses redshifted line emission (e.g.,\ 21-cm, Ly$\alpha$, CO, or {C}~\textsc{ii}) as a tracer for the underlying density field. Large volumes along a given line-of-sight may be surveyed simultaneously with a single spectrometer at relatively low spatial resolution, and by scanning this spectrometer across the sky a full 3D density map may be derived. Despite multiple different modeling efforts
\citep{Righi2008,Visbal+2010,Lidz+2011,Pullen+2013, Breysse+2014,Li+2016,Padmanabhan+2018, Moradinezhad+2019,Sun+2019,Yang2021,Moradinezhad2021,es_V} and significant progress on the observational front \citep{COPSS2,COLDz,mMIME,Keenan2021}, the overall level of the CO signal, especially in the clustering regime, is still unknown. 

The CO Mapping Array Project (COMAP; \citealp{es_I}) is an intensity mapping experiment that aims to use emission from carbon monoxide (CO) to trace the aggregate properties of galaxies over cosmic time, back to the Epoch of Reionization. A Pathfinder experiment, consisting of a 19-feed 26--34\,GHz receiver, has been fielded on a 10.4\,m single-dish telescope at the Owens Valley Radio Observatory (OVRO). In this frequency range, the receiver is sensitive to CO(1--0) at $z = 2.4$--3.4, with a fainter contribution from CO(2--1) at $z=6$--8. The main goal of the Pathfinder is to detect the CO(1--0) signal and use it to constrain the properties of galaxies at the Epoch of Galaxy Assembly. A future phase will add a second receiver at 12--20\,GHz in order to detect CO(1--0) from around $z=5$--9, cross-correlating with the CO(2--1) signal from the 26--34\,GHz receiver and constraining the properties of galaxies towards the end of the Epoch of Reionization.

The receiver's detector chain is based on cryogenically cooled HEMT low-noise amplifiers (LNA) which contribute to a typical system temperature of about 44\,K across the full frequency range. The predicted signal from high-redshift CO emission is expected to be no more than a few microkelvin per COMAP spatial/spectral resolution element (or ``voxel''). Thus, the raw instrumental noise must be reduced by many orders of magnitude before a statistically significant detection may be achieved. In practice, this is done by repeatedly observing the same part of the sky using multiple detectors, and thereby gradually increasing the sensitivity per voxel. For this to succeed, however, it is necessary to suppress systematic contributions from atmospheric temperature variations, sidelobe contamination, ground pickup, standing waves, Galactic foregrounds, etc.\ by a corresponding amount. 

The first season COMAP science observations started in June 2019 and lasted until August 2020. 
This paper describes the first season COMAP data analysis pipeline, which aims to produce clean maps from raw time-ordered COMAP observations. This includes calibration, data selection, filtering, and map-making. 
The rest of this paper is organized as follows: First, in order to establish useful notation and conventions, we give a brief introduction to the COMAP instrument in Section~\ref{sec:telescope}, while referring the interested reader to \citet{es_II} for full details. Next, we provide a high-level overview of the analysis pipeline in Section~\ref{sec:overview}, before specifying each step in Sections~\ref{sec:filter}--\ref{sec:mapmaking}. Data selection and efficiency is discussed in Sections~\ref{sec:selection} and~\ref{sec:transfer}. The results are presented in Section~\ref{sec:results}, and we summarize and conclude in Section~\ref{sec:conclusion}.

\section{Instrument and data model}\label{sec:telescope}

Before describing the COMAP analysis pipeline, we provide a brief overview of the instrument itself, and define an explicit data model. A more detailed description of the instrument can be found in a separate paper \citep{es_II}.

\subsection{Instrument overview}

The COMAP Phase I instrument observes in the ${\rm K_{a}}$ band, at 26--34\,GHz and is located at the Owens Valley Radio Observatory (OVRO) in California, USA. It is mounted on a 10.4~m telescope that was originally built for the Millimeter Array at OVRO, then used as a part of the Combined Array for Research in Millimeter-wave Astronomy (CARMA) experiment, and has now been repurposed for COMAP. The telescope's primary and secondary reflectors have diameters of 10.4\,m and 1.1\,m, respectively, and the beam FWHM is about 4.5 arcmin at 30\,GHz.

The receiver comprises 19 independent detector chains, called ``feeds''.
The signal chain of each feed consists of individual feed horns, polarizers, low noise amplifiers, two stages of downconversion, frequency separation and digitization. For the observations described in this paper, 15 feeds have a two-stage polarizer, two feeds have a single-stage polarizer, and two feeds have no polarizer. The digitization happens in two CASPER ``ROACH-2'' FPGA-based spectrometers for each signal chain, giving us four 2~GHz-wide sidebands (SB), each of which has 1024 frequency channels, resulting in a native frequency resolution of approximately 2\,MHz. The two sidebands of each band (A and B) are labelled  ``lower'' (LSB) or ``upper'' (USB). For more details on the instrument see \citet{es_II}. 

To support  frequent and accurate gain estimation, COMAP employs an ambient temperature load that is directly attached to the environmental shroud housing. This ``calibration vane'' is automatically moved in front of the feed horn array at the beginning and end of each observation (each lasting for about one hour; see Section~\ref{sec:observation}), fully filling the field of view of each pixel. The temperature of the calibration vane is monitored with sensors, allowing the system temperature to be calculated and applied to calibrate the gain (see Section~\ref{sec:gain} for more details).

\begin{table*}[htb!]
    \centering
        \caption{COMAP fields and calibrators} 
\begin{tabular}{c c c l}
        \hline
         Field Name & RA (J2000) & Dec (J2000) & Notes  \\
         \hline
         Field 1 & 01:41:44.4 & +00:00:00.0 & CO science field - lies within the HETDEX Fall field \\
         Field 2 & 11:20:00.0 & +52:30:00.0 & CO science field - lies within the HETDEX Spring field \\
         Field 3 & 15:04:00.0 & +55:00:00.0  & CO science field \\
         TauA & 05:34:31.9 & +22:00:52.2  & Pointing calibrator - supernova remnant (Crab Nebula) \\
         CasA & 23:23:24.0 & +58:48:54.0  & Pointing calibrator - supernova remnant \\
         CygA & 19:59:28.4 & +40:44:02.1  & Pointing calibrator - radio galaxy \\
         Jupiter &  &   & Pointing calibrator\\
         \hline
         \end{tabular}
    \label{tab:patch}
\end{table*}

\begin{figure}[t]
    \includegraphics[width=\linewidth]{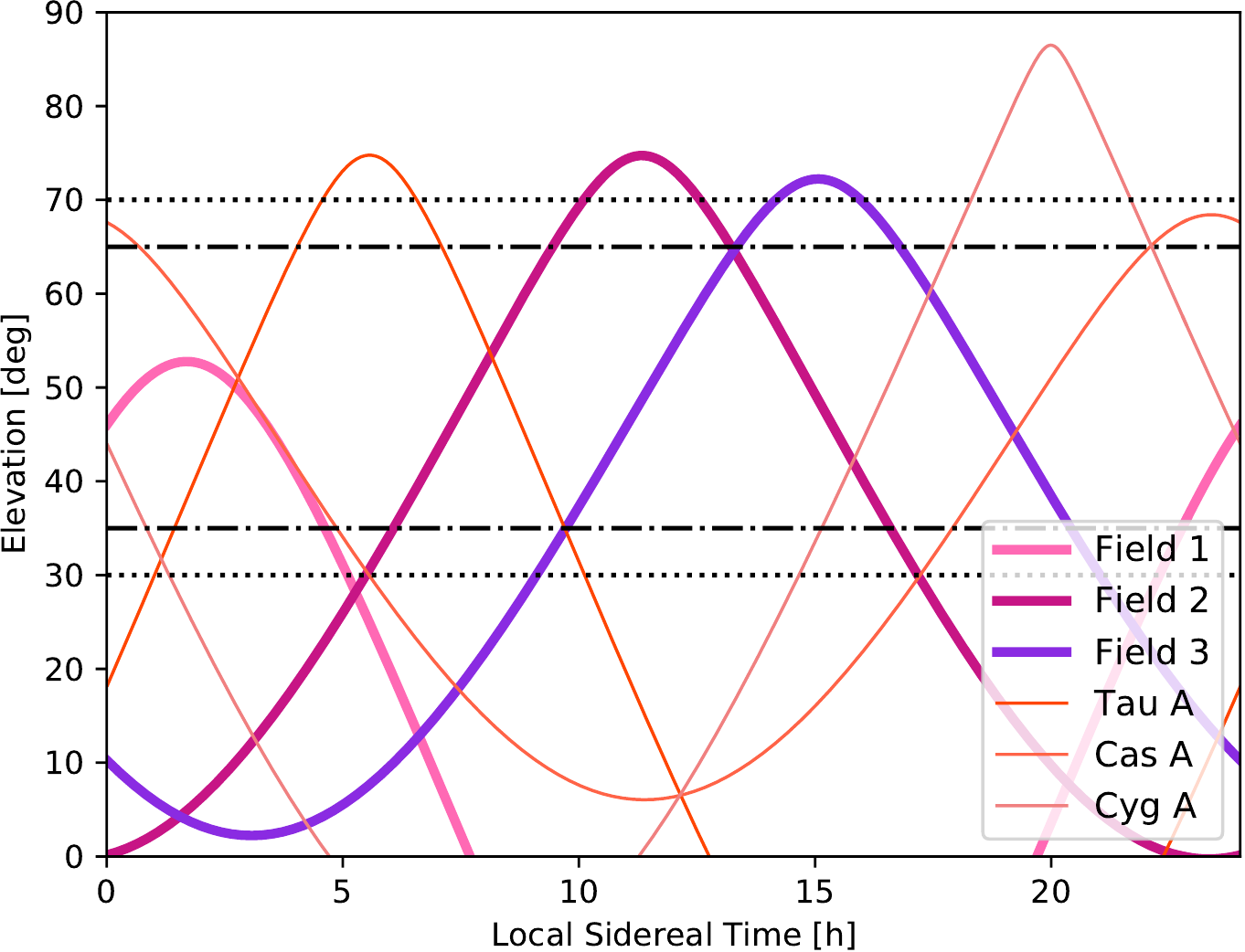}
    \caption{Elevation of CO (pink/purple) and calibration (orange) fields as a function of Local Sidereal Time. \label{fig:patches}}
\end{figure}

\begin{figure*}
  \includegraphics[trim=7.5cm 2cm 6.5cm 2.5cm, clip, width=\linewidth]{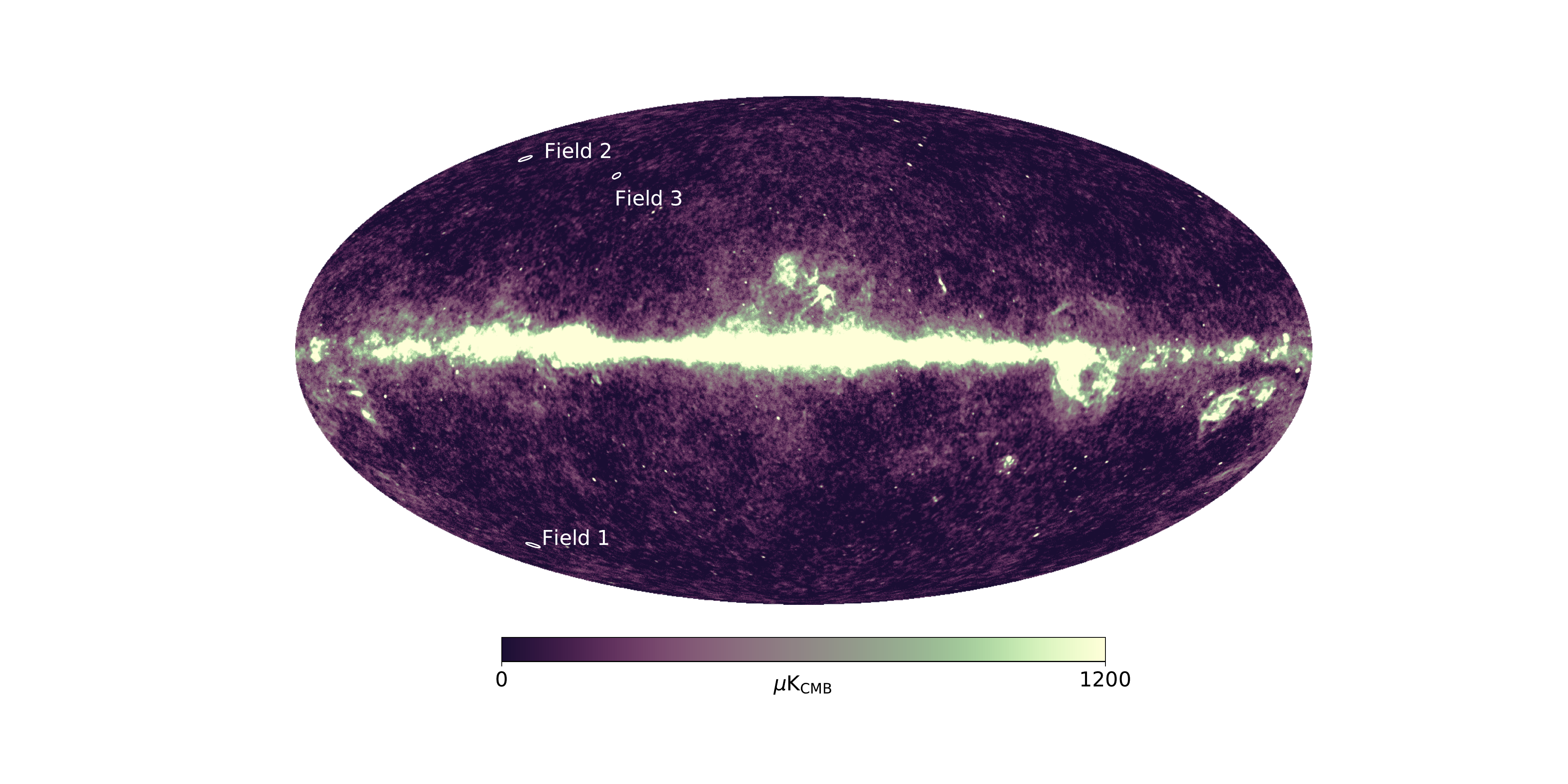}
  \caption{The three CO fields observed by the telescope overplotted as contours with radii of $\sim 1^\circ$, centered at the field centers (in Galactic coordinates) $(\mathrm{lon}, \mathrm{lat}) = (149.0^\circ, -60.3^\circ), (150.64^\circ, 59.53^\circ)$ and $(91.35^\circ, 53.22^\circ)$ for Fields 1, 2 and 3 respectively, on top of the \textit{Planck} LFI 30 GHz full-mission map (downloaded from the \textit{Planck Legacy Archive}
    \citealp{Planck2020}).}\label{fig:healpixfields}
\end{figure*}

\subsection{Field Selection}\label{sec:fields}
\begin{figure*}[t!]
    \includegraphics[ width=\linewidth]{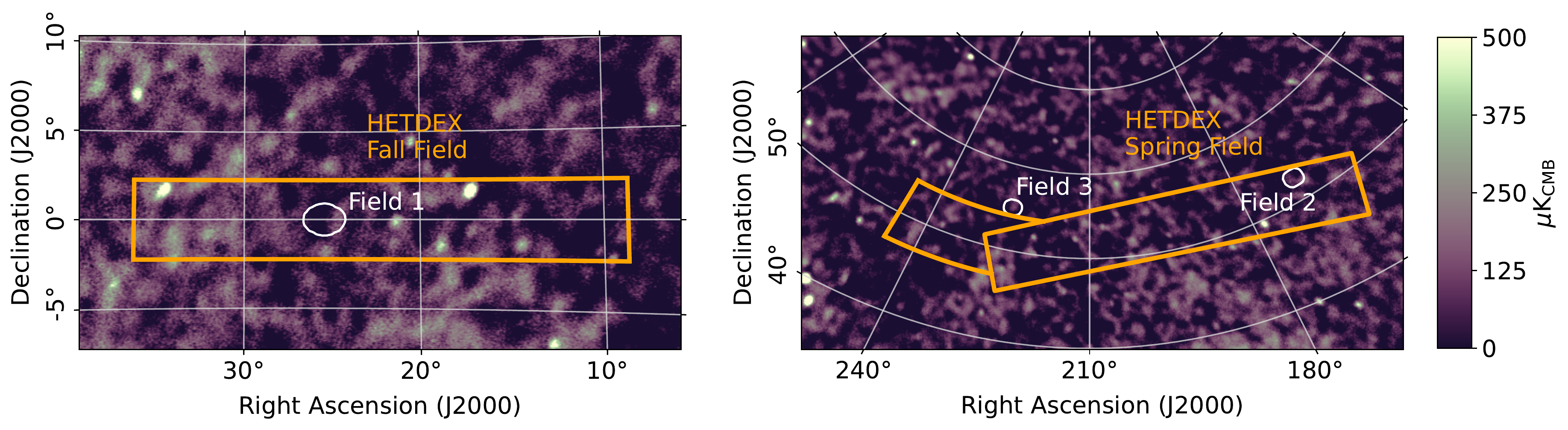}
    \caption{The three CO fields observed. The contours, illustrating the rough coverage of each field, have radii of $\sim 1^\circ$. In the left and right panels respectively we have drawn in the approximate coverage of the HETDEX Fall and Spring fields presented by \cite{Gebhardt+2021}. The map in the background is the same \textit{Planck} LFI 30 GHz full-mission map (downloaded from the PLA,  \citealp{Planck2020}) as seen in Figure \ref{fig:healpixfields}. \label{fig:cutoutfields}}
\end{figure*}

COMAP observes several parts of the sky. Table~\ref{tab:patch} lists all CO science fields and calibrators\footnote{Since COMAP began observing, the boundaries of the HETDEX Spring field coverage changed, with the result that one COMAP field no longer overlaps with the main HETDEX survey although we hope to also fill in this field with additional HETDEX observations.}. In Figure~\ref{fig:patches} we plot the elevation of the CO and calibration fields as a function of Local Sidereal Time, indicating when the fields are available for observation.
Figures \ref{fig:healpixfields} and \ref{fig:cutoutfields} show the position of the three CO fields observed by COMAP. These were selected to maximize the observing efficiency, avoid bright 30\,GHz point sources ($\gtrsim$ 1 Jy), and overlap with the coverage of Hobby-Eberly Telescope Dark Energy eXperiment (HETDEX; \citealp{Hill+2008,Gebhardt+2021,hill_etal_21}), a galaxy survey targeting Ly-alpha emission from galaxies in the same redshift.
Although COMAP's observing strategy has been designed to permit the direct detection of CO fluctuations from galaxies at $z=2.4-3.4$, cross-correlation with a galaxy survey such as HETDEX can increase the detection significance by at least a factor of two \citep{Chung+2019,silva_etal_21} as well as provide validation for the origin of detected signal in galaxies at the target redshift.

In addition to the main science fields, 
we are also conducting a survey of the Galactic plane covering longitudes $20^{\circ} < l < 220^{\circ}$, details of which can be found in \citet{es_VI}.

To facilitate calibration with astrophysical sources, we observe a handful of radio sources, including Jupiter, the supernova remnants Taurus A (TauA) and Cassiopeia A (CasA), and the radio galaxy Cygnus A (CygA), all of which are somewhat extended compared to the beam except for Jupiter.

\begin{figure*}
    \centering
    \includegraphics[width=1.0\linewidth]{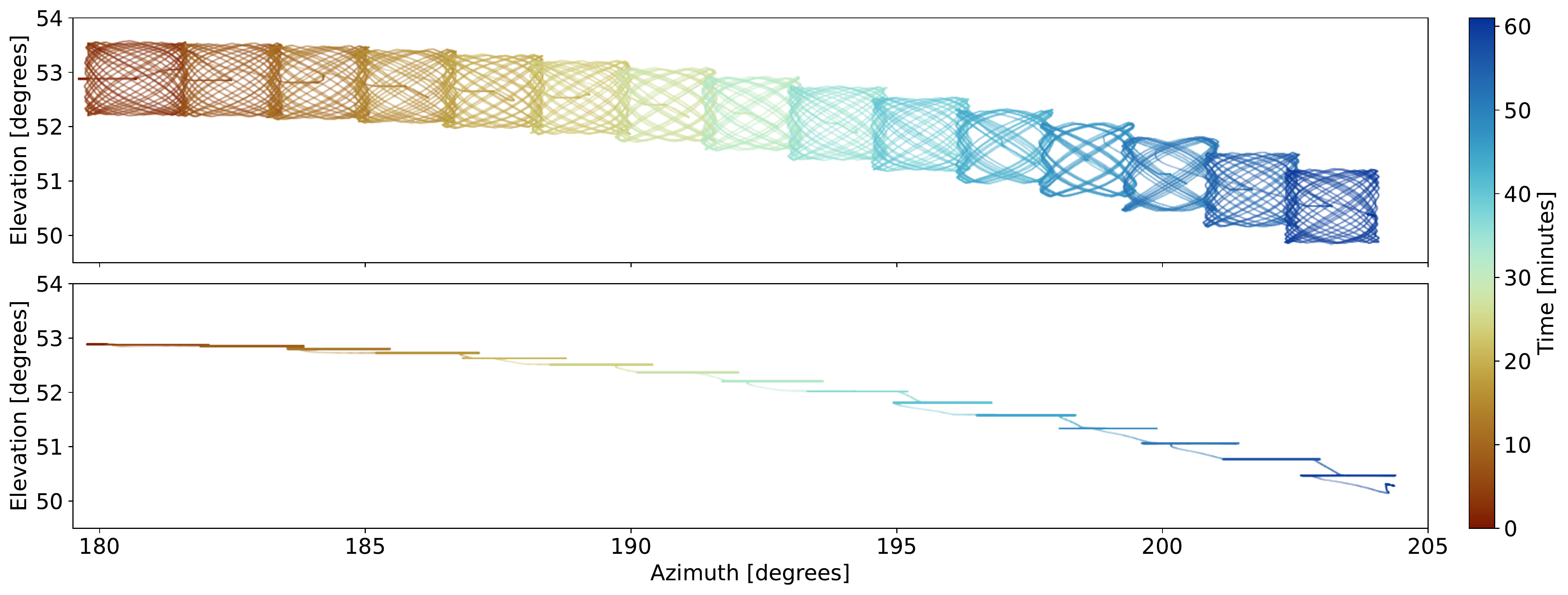}
    \caption{Movement of the telescope boresight in azimuth and elevation for an observation employing Lissajous scans (top) and an observation employing CES (bottom). Both observations consist of 15 individual scans of Field 1.
    \label{fig:Lissajous}}
\end{figure*}
\subsection{Observation Strategy}\label{sec:observation}
Telescope scans of the science fields follow a harmonic motion described by
\begin{equation}
    \mathrm{az} = A\sin(at + \phi);\,\,\,\,\, \mathrm{el} = B\sin(bt),
\end{equation}
where $A,B$ are amplitude parameters that define the size of the field, the ratio $a/b$ determines the shape of the curve, and $\phi$ is a phase parameter. Two different scan types were used: ``constant elevation scans (CES)'' ($b=0$) and ``Lissajous'' (varying parameters), alternating between each on a daily basis. At the start of a scan, the telescope is positioned at the leading edge of the field. The telescope then executes the scan while the field drifts through the pattern. This typically takes 3--10 minutes, after which the telescope is repointed to the leading edge of the field again in preparation for the next scan. An example of the scanning path for about one hour of continuous observations with a Lissajous scan and a CES is shown in Figure~\ref{fig:Lissajous}. Testing the relative performance of the CES and Lissajous scanning strategies in terms of final data quality is an important goal of the first-season COMAP survey.

\subsection{Data model}\label{sec:model}

As described by \citet{es_II}, the COMAP detector readout for a single frequency channel may be modelled as
\begin{equation}\label{eq:PTsys}
	P_\mathrm{out} = k_\mathrm{B} G \Delta\nu T_\mathrm{sys},
\end{equation}
where $k_\mathrm{B}$ is the Boltzmann constant, $G$ is the gain, $\Delta\nu$ is the bandwidth, and $T_\mathrm{sys}$ is the system temperature of the instrument. The system temperature may be further modeled as\footnote{In this section we are writing all the contributions to $T_\mathrm{sys}$ in terms of their effective noise contribution, rather than any physical temperatures. See Section~\ref{sec:gain} for a definition of $T_\mathrm{sys}$ in terms of physical quantities.}
\begin{align}
T_\mathrm{sys} =\,\, &T_\mathrm{receiver} + T_\mathrm{atmosphere} + T_\mathrm{ground} \nonumber\\
&+ T_\mathrm{CMB} + T_\mathrm{foregrounds} + T_\mathrm{CO},
\label{eq:Tsys}
\end{align}
where $T_\mathrm{reciever}$ is the effective noise temperature of the receiver, $T_\mathrm{atmosphere}$ is the noise contribution from the atmosphere, $T_\mathrm{ground}$ is ground pickup from far sidelobes, $T_\mathrm{CMB}$ is the contribution from the CMB, $T_\mathrm{foregrounds}$ are continuum foregrounds (typically from the galaxy), and $T_\mathrm{CO}$ is the line emission signal from extragalactic CO, which is the main scientific target of the COMAP instrument. 

To understand the challenges involved in measuring the cosmological CO signal, it is instructive to consider the order of magnitude and stability of each term in Equation~(\ref{eq:Tsys}). The largest single contribution is that of the receiver temperature, which is usually about 10--30\,K. For the COMAP receiver, with HEMT LNA technology, this is very stable. 

The second-largest contribution is from the atmosphere, which typically adds 15--25\,K. This term varies significantly on all time scales longer than a few seconds, and depends on external conditions including elevation, humidity, cloud coverage, ambient temperature and wind speed. It is also strongly correlated between detectors and frequencies, since all feeds observe through essentially the same atmospheric column at any given time; fortunately, the phase structures of the atmospheric fluctuations are uncorrelated on long time scales. 

Next, ground pickup typically accounts for 5--6\,K, and this term can be particularly problematic because it depends sensitively on the instrument pointing: If a sidelobe happens to straddle a strong signal gradient, such as the horizon or the Sun, several mK variations may be measured on very short timescales and with a time-dependency that appears nearly sky synchronous. 

The fourth term represents the CMB temperature of 2.7\,K, which is both isotropic and stationary, while the fifth term represents astrophysical foregrounds, expected to contribute at most 1\,mK; for instance synchrotron, free-free, and dust emission from the Galaxy. Although these are sky synchronous, and in principle could confuse potential CO measurements, they also have very smooth frequency spectra \citep{Keating+2015}, and are therefore relatively easy to distinguish from the cosmological CO signal, which varies rapidly with frequency. An important potential exception is line emission from other molecules redshifted to our band from galaxies at other epochs. The hydrogen cyanide (HCN) line is expected to be one of the brightest such lines. Emission from HCN in galaxies towards our CO fields at redshift $z = 1.6$--2.4 will appear in our frequency range. However, this contribution is expected to be an order of magnitude lower than that from CO \citep{Chung+2017}. 

Finally, the cosmological CO line emission signal is expected to account for $\mathcal{O}(1\,\mu\textrm{K})$. Whether it is possible to detect such a weak signal depends directly on the stability and sensitivity of the instrument. In this respect, the fundamental quantity of interest is the overall noise level of the experiment, which is dominated by random thermal noise. 

The magnitude of these random thermal fluctuations is proportional to  $T_\mathrm{sys}$, with a standard deviation that is given by the so-called radiometer equation,
\begin{equation}\label{eq:radiometer}
	\sigma_\mathrm{N} = \frac{T_\mathrm{sys}}{\sqrt{\Delta \nu\,\tau}},
\end{equation}
where $\tau$ is the integration time. Thus, since both the system temperature and the bandwidth are essentially fixed experimental parameters, the only way of reducing the total uncertainty is by increasing the integration time. As a concrete and relevant example, we note that an integration time of 45 hours is required to achieve a standard deviation of $20\,\mu\mathrm{K}$ with a system temperature of 45\,K and a bandwidth of 31.25\,MHz. 

\begin{figure*}[t]
    \includegraphics[width=\linewidth]{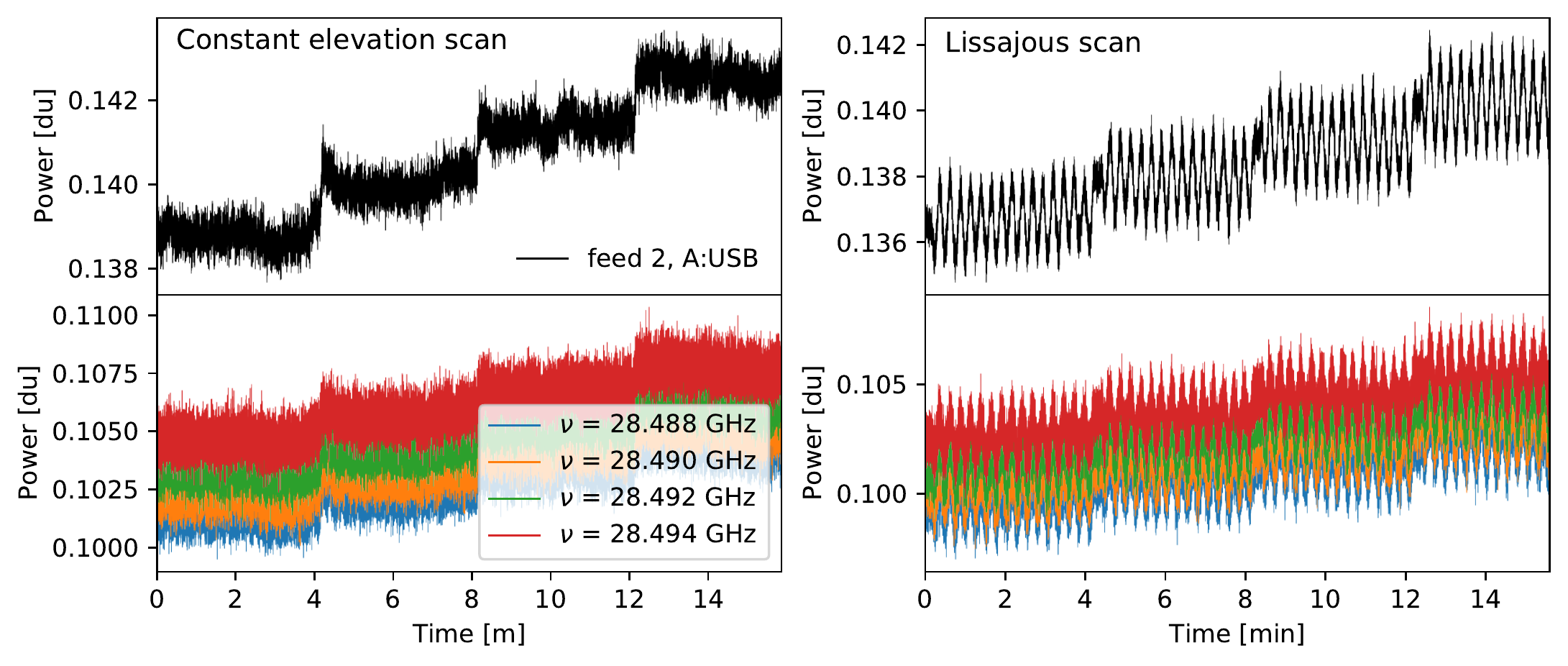}
    \caption{Raw data from the COMAP instrument (in arbitrary digital units of power)
      Here we see data averaged over a single 2\,GHz-wide sideband (top) and examples of data from four individual frequency channels in that sideband (bottom). These data were taken using two different scan patterns: CES (left) and Lissajous (right).
    \label{fig:raw_tod}}
\end{figure*}

\begin{figure*}[t]
    \includegraphics[width=\linewidth]{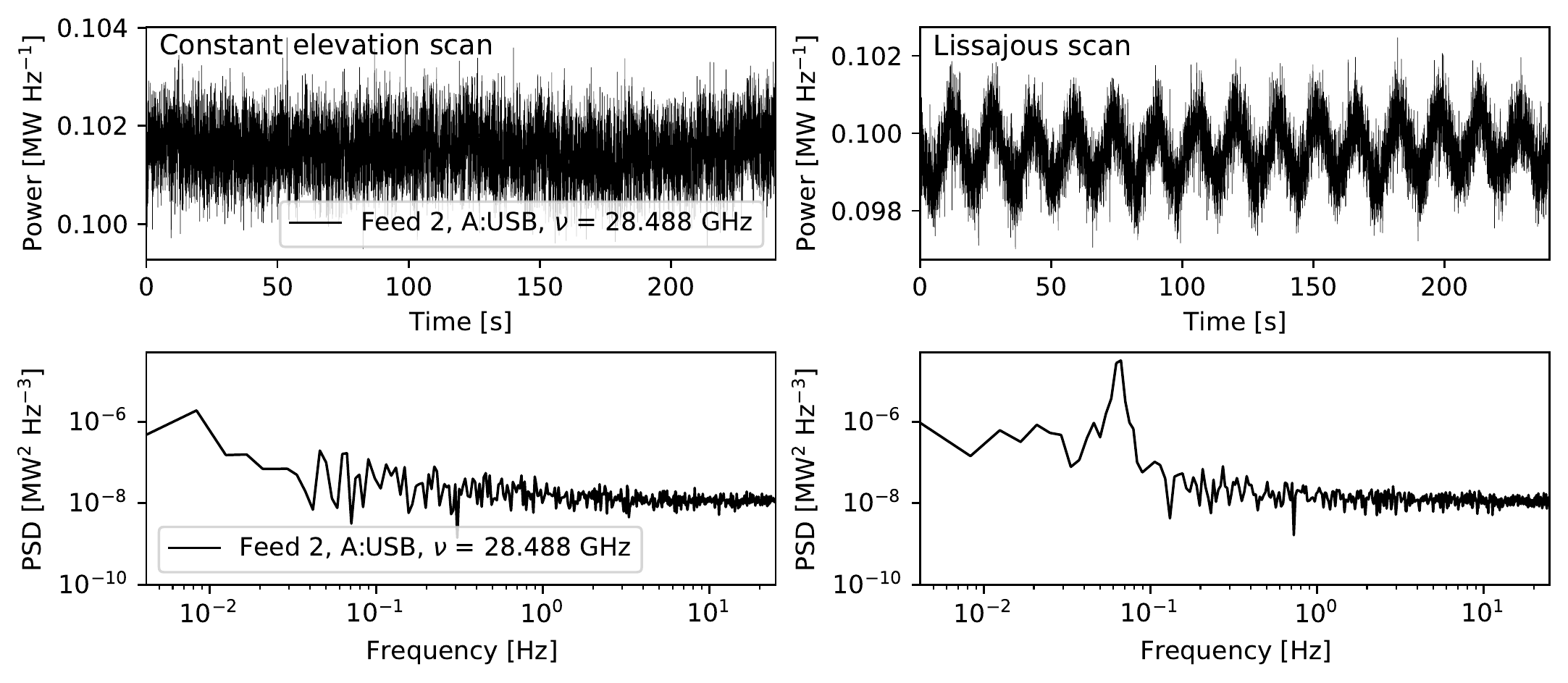}

    \caption{Raw data from an individual frequency channel of the COMAP instrument. Power is shown as a function of time (top), and the corresponding power spectral density (PSD) is also shown (bottom). We show data from a CES scan (left) and a Lissajous scan (right).
    \label{fig:tod_ps}}
\end{figure*}

\begin{figure*}[t]
    \includegraphics[width=\linewidth]{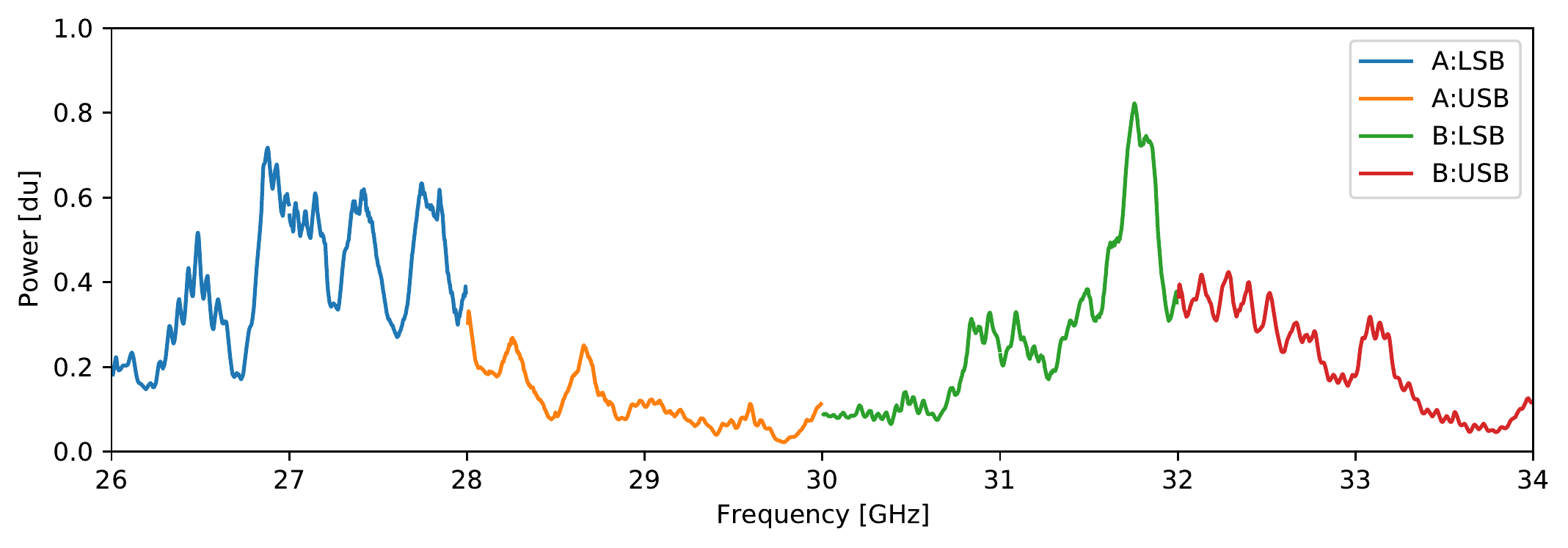}
    \caption{Time-averaged raw data from each frequency channel on a single feed of the COMAP instrument. The colors represent the four 2\,GHz-wide sidebands. Note that a few of the frequency channels at at the edges and middle of sidebands tend to be unstable and are masked out in the analysis. 
    \label{fig:freq_spectrum}}
\end{figure*}
\begin{figure}[t]
    \includegraphics[width=\linewidth]{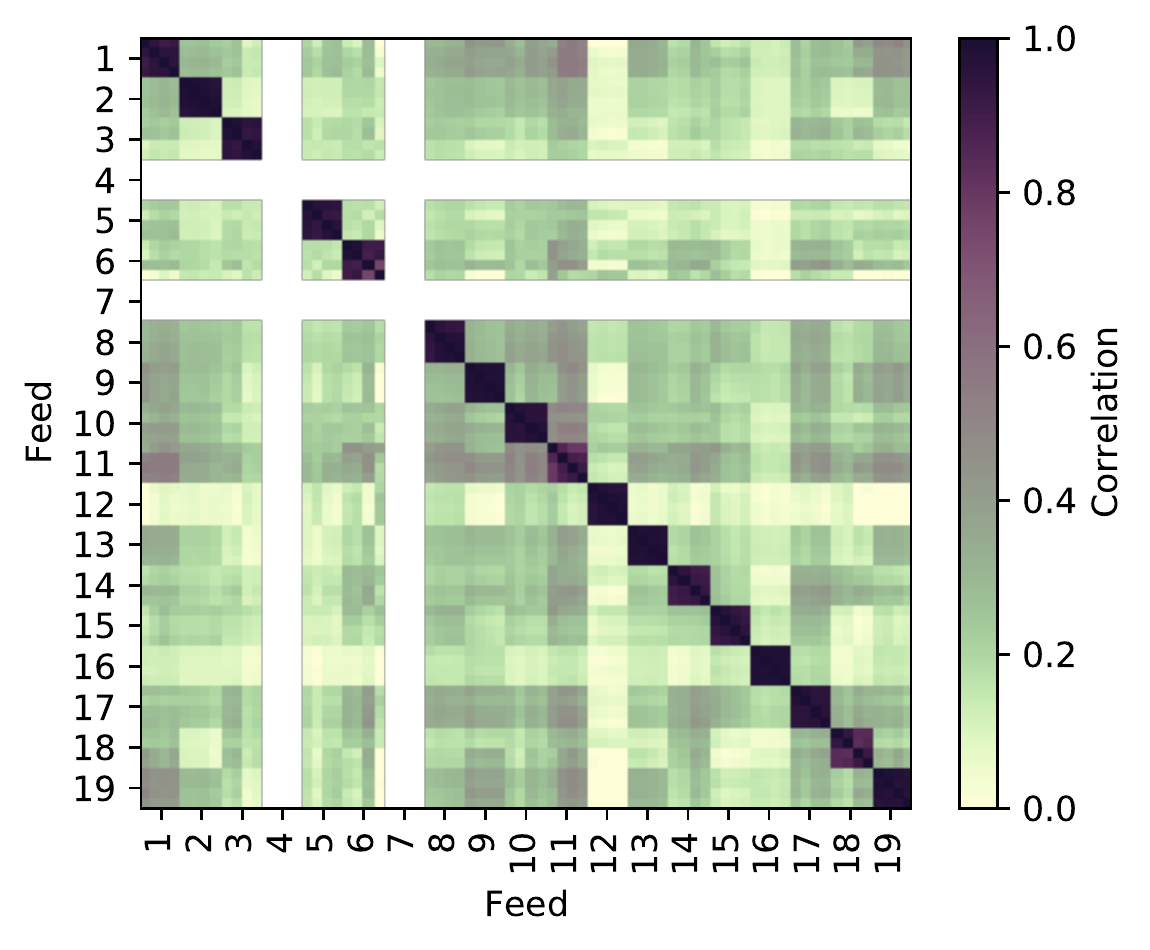}
    \caption{Correlation between the sideband-averaged data from the 19 feeds of the COMAP instrument for a single constant elevation scan. For this observation, as for much of the observing campaign, the LNAs for feeds 4 and 7 were turned off because those feeds, as a test, did not have a polarizer and so had large standing waves due to reflections between the receiver and the secondary reflector. 
    \label{fig:sb_mean_corr}}
\end{figure}

\begin{figure}[t]
    \includegraphics[width=\linewidth]{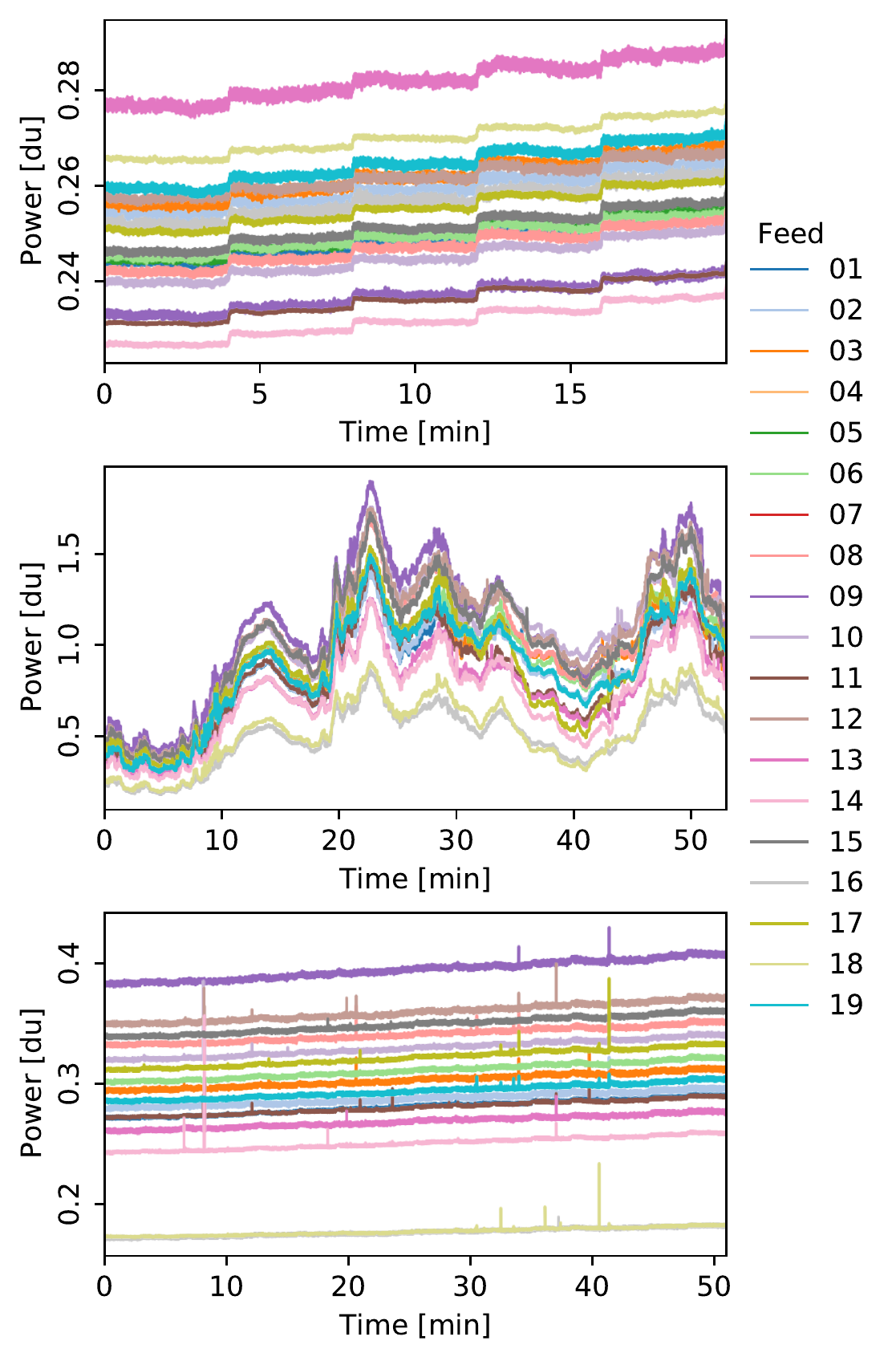}

    \caption{Feed averaged COMAP TOD recorded under various observing conditions. The top panel shows data observed under normal conditions, and is dominated by instrumental noise. The middle panel shows data observed under poor weather conditions with a thick cloud coverage, resulting in large coherent power fluctuations observed by all feeds. This third panel shows data with strong spikes, which may for instance happen during rare periods with high insect activity. 
    \label{fig:all_feed_comparison}}
\end{figure}

\begin{figure}
    \centering
    \includegraphics[width=\linewidth]{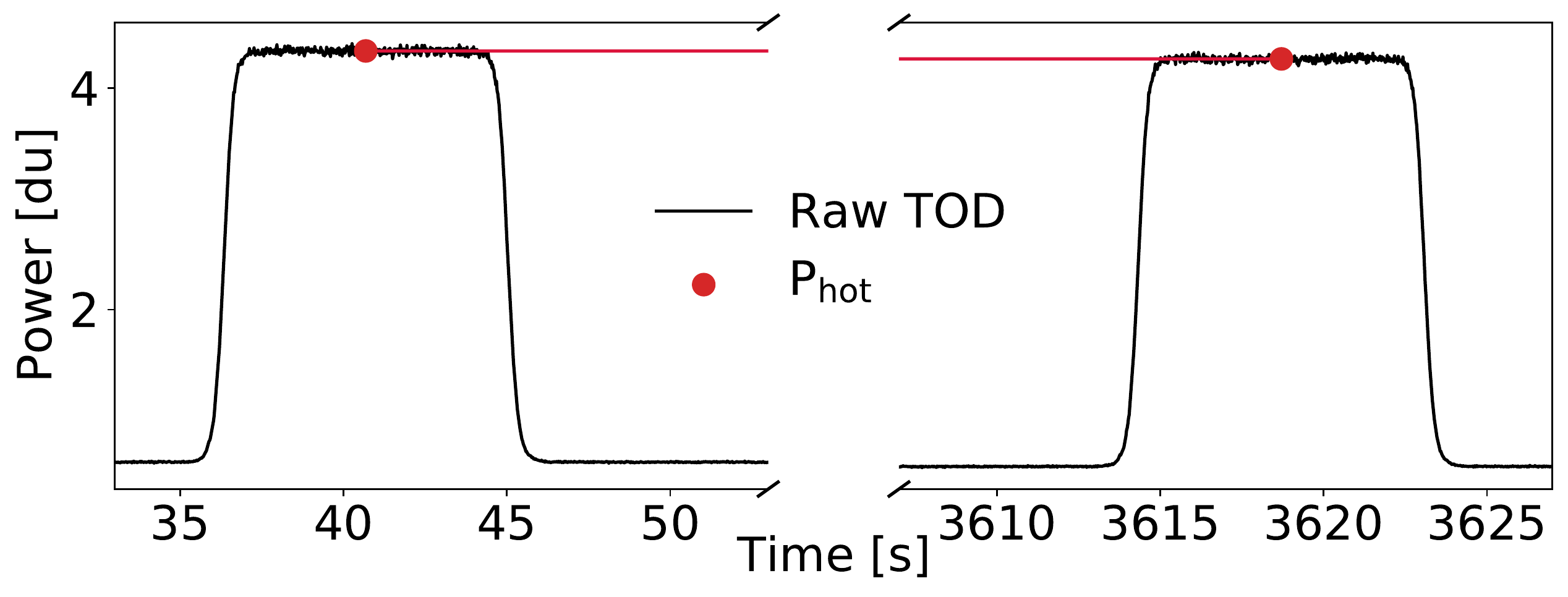}
    \caption{The calibration vane is inserted in front of the receiver at the beginning and end of one observation of a CO science field. The time between calibration vane insertions is typically about an hour, a period set by the preferred data file size for the CO field observations.}
    \label{fig:TOD_Tsys_calib}
\end{figure}

In addition to the thermal and uncorrelated noise described by the radiometer equation, there are three main sources of correlated noise, namely gain fluctuations in the low-noise amplifiers, atmospheric temperature fluctuations, and time-dependent standing waves. All of these are expected to have a roughly $1/f$-type spectrum, although with different particular properties\footnote{There are several different sources of standing waves, some of the main ones give rise to $1/f$-like spectra, but others do not.}. The fact that these sources of correlated noise are also strongly correlated between frequencies is very useful in order to filter out this noise in the analysis.

Equation~(\ref{eq:PTsys}) describes the detector output at any given time. To connect this to the actual measurements recorded by the detector, we adopt the following data model,
\begin{align}
    d_\nu^i(t) &= \langle d_\nu^i \rangle(1 + \delta^i_G(t)) \left[1 + P^i_\mathrm{cel}(\Delta s_\mathrm{cont} + \Delta s^\nu_\mathrm{CO}) \right. \nonumber\\
    &+ \left.P^i_\mathrm{tel}\Delta s_\mathrm{ground} + n_\mathrm{corr}(t)  + n^{\nu i}_\mathrm{w}(t)\right].
\end{align}
Here $d_\nu^i(t)$ denotes the raw data recorded at time $t$ for frequency channel $\nu$ in feed $i$; $\langle d_\nu^i \rangle$ represents the corresponding time average and basically corresponds to $\langle {T^{i\nu}_\mathrm{sys}(t)} \rangle \langle G^i_\nu(t) \rangle$; $\delta_G^i(t)$ denotes feed dependent gain fluctuations; $P^i_\mathrm{cel}$ and $P^i_\mathrm{tel}$ are pointing matrices in celestial and telescope coordinate systems, respectively; $\Delta s_\mathrm{cont}$ denotes the celestial continuum source fluctuations, mainly from the CMB and Galactic foregrounds; $\Delta s^\nu_\mathrm{CO}$ is the CO line emission fluctuation; $\Delta s_\mathrm{ground}$ is the ground signal fluctuation picked up by the far sidelobes; and $n_\mathrm{corr}(t)$ are the correlated temperature fluctuations, mostly consisting of atmosphere fluctuations and standing waves. Factors with no feed or frequency index are assumed to be similar (or at least strongly correlated) at different frequencies and feeds, while factors with a $\nu$ label indicate parts of the model that are assumed to have non-smooth frequency dependence. The main purpose of the COMAP analysis pipeline is to characterize $\Delta s^\nu_\mathrm{CO}$ given $d_\nu^i(t)$.

\subsection{Data overview}\label{sec:data_overview}

Before presenting the analysis pipeline, we provide a preview of the raw time-ordered data (TOD) generated by the COMAP instrument, with the goal of building intuition that will be useful for understanding the purpose of each component of the analysis pipeline described in this paper. 
Figures~\ref{fig:raw_tod} and \ref{fig:tod_ps} show examples of such raw time-ordered data (TOD) from the instrument using the CES (left column) and Lissajous (right column) scanning strategies. Perhaps the most obvious features in these plots are step-wise changes in power as the telescope changes elevation during repointings between scans; see Section~\ref{sec:observation}. The Lissajous scans additionally show oscillations in power as the telescope changes elevation during the scan, since the telescope looks through a thicker slab of atmosphere at lower elevations, and this increases the atmospheric contribution to the system temperature. 

The top panels in Figure~\ref{fig:tod_ps} show an individual frequency channel for a single scan (i.e., stationary observation period), while the bottom panel shows the corresponding power spectral density (PSD). For the CES case, the PSD is relatively featureless, with an overall shape that looks consistent with a typical $1/f$ noise spectrum. For the Lissajous case, an additional strong peak is seen around 0.007\,Hz, which matches the scanning period of 14\,sec, and this corresponds to the periodic atmospheric variations seen in the panels above.

Figure~\ref{fig:freq_spectrum} shows the time averaged data for all frequency channels of a single feed for one scan. The spectral shape is mostly determined by the average gain as a function of frequency, due to the combined effect of the various components of the receiver chain. This average gain is a purely instrumental effect, not associated with the true sky signal, and therefore simply corresponds to a normalization factor that should be calibrated out before higher-level analysis. However, some of the spectral shape is also determined by the fact that the system temperature also changes with frequency, and in some cases exhibits large spikes within specific frequency ranges (see \citet{es_II} for more details). Separating the gain variation as a function of frequency from the system temperature as a function of frequency is a main goal of the calibration procedures described in Section~\ref{sec:gain}. 

In Figure~\ref{fig:sb_mean_corr} we plot the correlation,
\begin{equation}
    C_{ij} = \frac{\left<\hat{d}^i \hat{d}^j\right>}{\sqrt{\left<\hat{d}^i\hat{d}^i\right>\left<\hat{d}^j\hat{d}^j\right>}},
\end{equation}
between the power, $\hat{d}^i$ recorded by any two feeds, $i$ and $j$,  after 
averaging over all frequencies within each sideband for each radiometer. Here we first note that the data from different sidebands of the same feed are strongly correlated. This is because both main sources of correlated noise in the COMAP data, namely gain fluctuations and atmospheric fluctuations, are common for sidebands within a given feed. In contrast, sidebands for different feeds mostly share the atmospheric fluctuations (and also some standing waves), but have independent gain fluctuations, and this results in lower overall correlations, but still typically in the 10--40\% range. Accounting for and mitigating such correlations will clearly be essential in order to extract robust science from these observations.

The quality of the COMAP data depends strongly on the observing conditions, as illustrated in Figure~\ref{fig:all_feed_comparison}. The top panel shows an observation made under normal conditions, while the middle panel shows an observation made during poor weather, with thick cloud coverage. The bottom panel shows a data segment with strong ``spikes'', a feature of some data taken in summer, possibly associated with insects flying in front of the focal plane. Automatic identification and removal of problematic data is clearly an important and necessary component of the pipeline.

Finally, Figure~\ref{fig:TOD_Tsys_calib} shows the calibration vane observations that are made at the beginning and end of each observation period. Since the ambient temperature is about one order of magnitude higher than $T_{\mathrm{sys}}$, the measured power is also correspondingly about one order of magnitude higher, and this bright and known signal allows for a precise  estimate of $T_\mathrm{sys}$. Note that these data segments are removed prior to data analysis, as they would otherwise compromise any filtering that may be applied to the data.  

\section{COMAP Analysis Pipeline}\label{sec:pipeline}
\subsection{Pipeline Overview}\label{sec:overview}

\begin{figure*}[t!]
  \includegraphics[width=\linewidth]{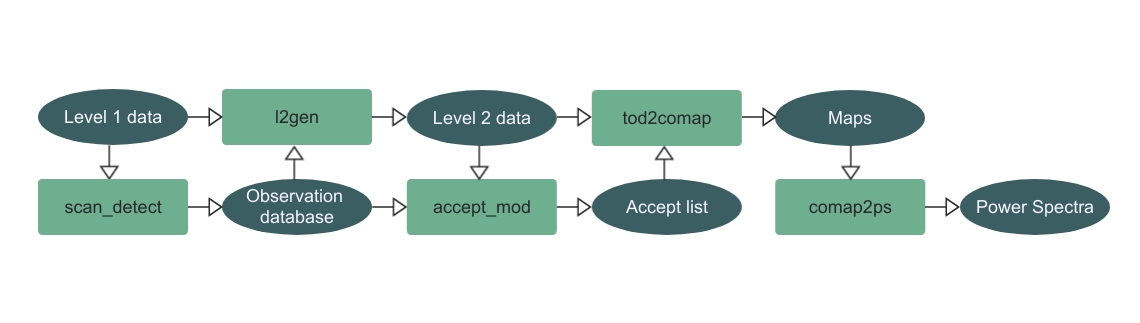} 
  \vspace*{-1.1cm}
    \caption{Flow diagram of the analysis pipeline. The dark green ellipses are data products, whereas the light green boxes are the different modules of the data analysis pipeline. \label{fig:flow}}
\end{figure*}

We are now ready to present the COMAP analysis pipeline, which is designed to process the raw data discussed in Section~\ref{sec:data_overview} into calibrated and cleaned CO maps. The main steps of this pipeline are schematically illustrated in Figure~\ref{fig:flow}.

The processing starts with ``Level 1'' files, which contain raw data as recorded by the instrument, together with pointing information and house-keeping data. Each of these files typically contain about one hour of observation time, including calibration vane observations at the beginning and end. 
We denote each (rougly) one hour of data as one observation, and assign it an individual observation ID (abbreviated obsID). Each observation consists of several scans, where one scan is the period between two re-pointings of the telescope, during which the telescope performs the same motions around a fixed point in azimuth and elevation while the target field drifts through. The instrumental properties are consequently assumed to be stationary within each scan. The module denoted \texttt{scan\_detect} in Figure~\ref{fig:flow} indicates a dedicated code that partitions each observation into individual scans, based on pointing information, and records information of each scan in a database. 

The main processing takes place in the \texttt{l2gen} module, which generates calibrated and cleaned TOD and stores them in so-called ``Level 2'' files. This is achieved through the application of a series of filters (see Section~\ref{sec:filter}) and a time-varying gain normalization (see Section~\ref{sec:gain}). This stage also evaluates basic goodness-of-fit statistics and defines a frequency channel mask that excludes missing or broken data for the current scan, before reducing the spectral resolution of the data to a spectral resolution suitable for map-making. In our main analysis, we reduce the resolution from $\sim$ 2\,MHz to $\sim$ 31\,MHz, resulting in the computational speed-up of subsequent steps and a memory saving for storing final maps by a factor of 16.

Next, the \texttt{accept\_mod} module reads in the statistics (including goodness-of-fit)  and basic frequency mask produced by \texttt{l2gen} and produces a list of accepted observations as defined by user-specified thresholds for each statistic (see Section~\ref{sec:selection}). Examples of relevant statistics used for this purpose are $\chi^2$ per observation, correlated noise knee-frequency ($f_{\mathrm{knee}}$), and Solar elongation. The output from this process is called an \emph{accept list}, which determines what data to use for mapmaking.

Converting time-ordered data into pixel-ordered data is done by a map-maker called \texttt{tod2comap} (see Section~\ref{sec:mapmaking}). As shown in the following sections, the adopted filters result in very nearly uncorrelated white noise, and the current implementation of \texttt{tod2comap} accordingly adopts simple binning into voxels. Finally, from these maps we can estimate the CO power spectrum using the module \texttt{comap2ps} (see \citealp{es_IV} for details).

\subsection{Data Segmentation}
As described above, we define a \emph{scan} to be the observing period between re-pointings of the telescope. 
The purpose of the \texttt{scan\_detect} code is to identify all scans within all observation periods, and produce an 
observation database, consisting of a list of obsIDs sorted according to source. For each obsID, we list all scans within that obsID, including basic information such as the modified Julian date (MJD) of the start and end of the scan, as well as the scanning mode (e.g. Lissajous or CES) and mean pointing information. 

\subsection{Filtering}\label{sec:filter}
As described in Section~\ref{sec:overview}, the COMAP TOD exhibit a wide range of non-CO-related contributions, both of instrumental and external origin. These must be suppressed by orders of magnitude prior to map-making in order to extract the astrophysically valuable signal. With this goal in mind, we introduce four specific filters, each targeting one class of artifacts.

\begin{figure*}
    \centering
    \includegraphics[width=\linewidth]{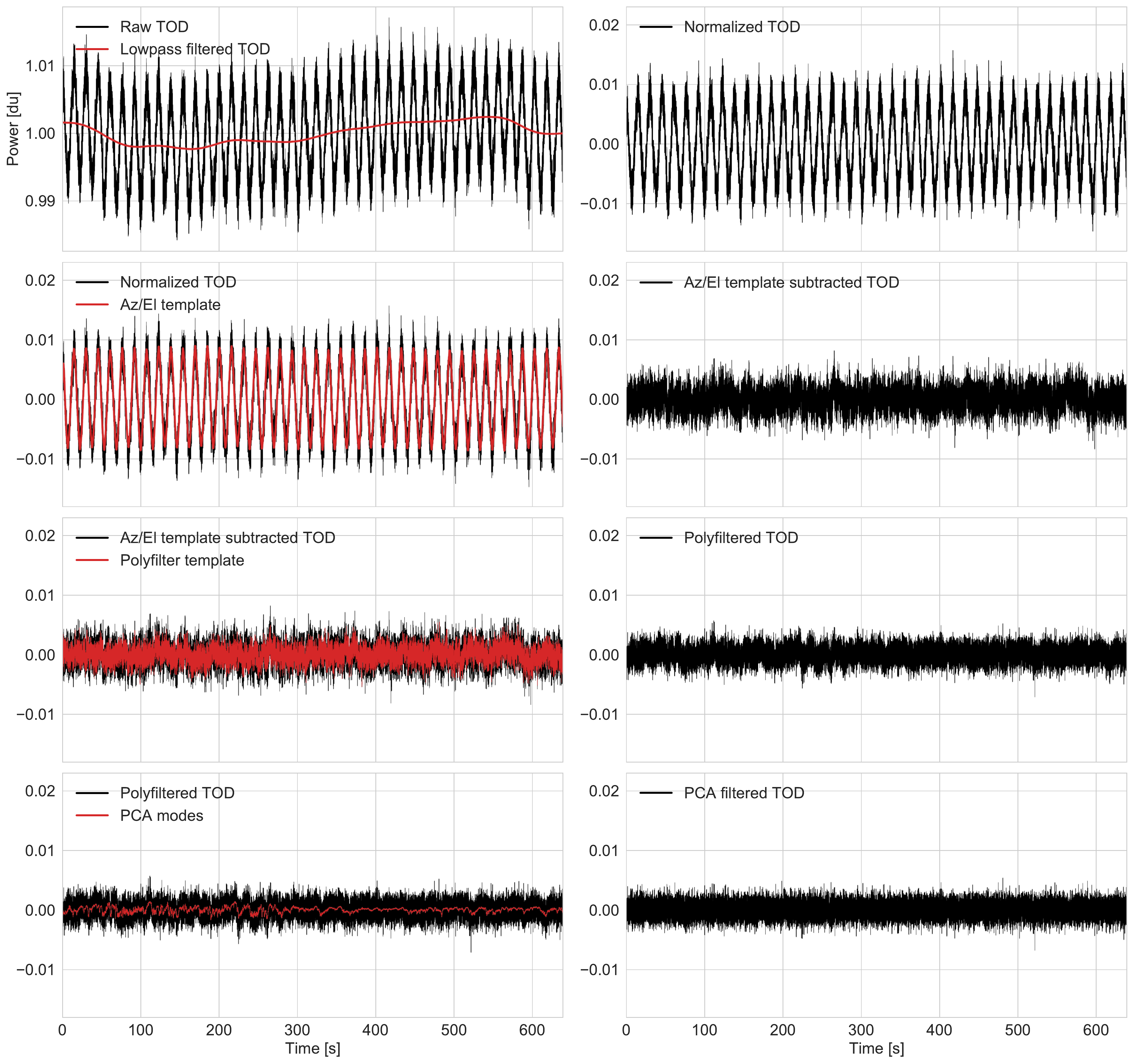}
    \caption{Effect of each filter in time-domain. Each row shows the data before (left column) and after (right column) applying the indicated filter. From top to bottom, the filters shown are 1) normalization; 2) elevation gain subtraction; 3) poly-filtering; and 4) PCA filtering. Data used is from scan 7717.03, feed 9, in a 31.25\,MHz band around 32.3\,GHz.}
    \label{fig:TOD_pipeline}
\end{figure*}

Figure~\ref{fig:TOD_pipeline} shows the evolution of the data as it passes through each of the filters.

\subsubsection{Normalization}

The first filtering operation we introduce is data normalization. This is done simply by dividing the raw TOD, $P_\mathrm{out}$, by its own running mean, and then subtracting one,
\begin{equation}
	d(\nu,t) = \frac{P_\mathrm{out}(\nu,t)}{\langle P_\mathrm{out}(\nu,t) \rangle} -1.
	\label{eq:normalization}
\end{equation}
Here $t$ is a time sample index and $\nu$ denotes frequency channel. This operation is performed separately on each frequency channel. The running mean is estimated by putting the data through a lowpass filter with a times scale of about 100 seconds. This step basically removes $\langle d^i_\nu \rangle$ from Equation~(\ref{eq:data_model}), and it also removes the first term in the square bracket (which is equal to 1) of the same equation. 

The main purpose of this step is to equalize (i.e., ``flatten'') the instrumental passband, as illustrated in Figure~\ref{fig:freq_spectrum}, and effectively establish data with appropriate relative calibration. The main practical advantage of doing so is that the amplitude of common-mode contaminants, such as gain-induced correlated noise or atmospheric fluctuations, become comparable across all frequencies within a single sideband, and therefore much easier to filter out. The same also holds true for broadband astrophysical contributions, such as the CMB or foregrounds, which also must be removed prior to signal extraction. See top panel of Figure~\ref{fig:TOD_pipeline} to see the effect of the normalization step. We can see that long timescale fluctuations are removed, and that the data now fluctuates around zero. 

Note also that with the definition in Equation~(\ref{eq:normalization}), the noise level of $d(\nu,t)$ is given by the sample rate and bandwidth alone in the ideal case, and should equal $1/\sqrt{\tau\Delta\nu}$. Calibration into physical units is performed simply by multiplying $d(\nu,t)$ with $T_{\mathrm{sys}}$. We find that $d(\nu,t)$ is a particularly convenient function for goodness-of-fit tests and it will serve as our main object of interest in the following.

\subsubsection{Removal of Az/El Templates}

The second filter we apply is designed to suppress signals that are correlated with local pointing (azimuth and elevation), as opposed to sky-correlated signals. The two main effects of this type are elevation-correlated atmospheric contributions and azimuth-correlated sidelobe contributions. The first of these effects may be modelled by a simple expression for the optical depth of the atmosphere of the form
\begin{equation}
 \tau(\mathrm{el}) = \frac{\tau_0}{\sin(\mathrm{el})},
\end{equation}
where $\tau_0$ is the optical depth of the atmosphere at zenith, and $\mathrm{el}$ is the elevation, while the second effect may be approximated through a low-order polynomial in azimuth. We therefore filter the data by fitting and subtracting the following simple model to each normalized frequency channel separately,
\begin{equation}
 d = \frac{g}{\sin(\mathrm{el}(t))} + a \,\mathrm{az}(t) + c + n.
\end{equation}
Here $g$, $a$ and $c$ are fitting constants, and $n$ denotes Gaussian noise with an assumed constant variance. We find the best-fit values for the free parameters by minimizing a $\chi^2$ statistic, and use $g$ and $a$ to clean the TOD with respect to the Az/El templates, 
\begin{equation}
 d_\mathrm{after} = d_\mathrm{before} - \frac{g}{\sin(\mathrm{el})} - a \, \mathrm{az}  - \left\langle \frac{g}{\sin(\mathrm{el})} + a \,\mathrm{az} \right\rangle.
\end{equation}
In this expression, $\langle \ \rangle$ denotes the mean value in time for a specific frequency channel, and this term ensures that the TOD has vanishing mean also after subtraction of Az/El templates. For long-duration scans we divide the TOD into disjoint segments of roughly 4 minutes each, and perform the template fit and removal separately on each data segment, in order to improve the tracking of temporal variations.

The effect of the Az/El template removal can be seen in the second row of Figure~\ref{fig:TOD_pipeline}.

\subsubsection{Polynomial continuum filter}
\label{sec:poly}
Our third filter, usually referred to as the ``polyfilter", is designed to remove all continuum signals observed simultaneously by all frequency channels within a given sideband. Specifically, for each timestep we fit and subtract a low-order (and typically linear) polynomial to the normalized and Az/El-subtracted TOD in frequency space for each sideband. 

We assume
\begin{equation}\label{eq:polymodel}
 d_\nu = c_0 + c_1 \nu + c_2 \nu^2 + ... \ ,
\end{equation}
where $d_\nu$ are the data across one sideband at a specific timestep, $c_0, c_1$ and $c_2$, etc., are constants that are fitted independently for each sideband and at each time step. We then remove the fitted polynomial from the data. In the third row of Figure~\ref{fig:TOD_pipeline}, we can see an example of how this filter removes the majority of the correlated noise from the data. 

The main target of this filter is $1/f$ noise from gain variations in the receiver electronics and atmospheric temperature fluctuations, which is strongly correlated between frequency channels within each sideband. Indeed, the fact that this noise is so tightly correlated between channels is one of the key instrumental features of the COMAP instrument that makes CO measurements feasible in the first place, effectively reducing the final noise level by a significant amount. 

As a bonus, this polynomial filter also suppresses any slowly varying astrophysical signal, and in particular broadband signals such as CMB, synchrotron, free-free or anomalous microwave emission. In contrast, the cosmological CO signal is expected to vary on the scale of adjacent frequency channels, and is therefore only mildly affected by this filter. However, some CO signal is indeed lost on the largest longitudinal scales due to this filter, and this effect will later be quantified in terms of an effective transfer function (see Section~\ref{sec:transfer} for more details). 

\subsubsection{Principal Component Analysis (PCA) filter}

While the previous filter removes continuum signals within each sideband, our fourth and final filter targets common-mode signals seen simultaneously by the entire focal plane. The two most prominent examples of such contaminants are residual atmospheric variations and standing waves, both of which have strongly correlated time variations across all feeds and frequencies. To suppress these signals, we perform a so-called Principal Component Analysis (PCA) on the whole data set, and subtract the leading modes. Intuitively speaking, this amounts to identifying the functions of time that explain the largest amount of the variance between the different frequencies across all the different feeds. These functions are the leading PCA components.

To formulate this idea in a mathematical language, let us organize all data in given scan into a data matrix $D$, where each row contains the TOD corresponding to a single frequency channel on a single feed. Thus $D$ is a matrix with dimensions $n_\mathrm{freq} \times n_\mathrm{samp}$, where $n_\mathrm{freq}= n_\mathrm{feeds} \cdot n_\mathrm{sidebands} \cdot n_\textrm{freq per sideband} = 19\cdot 4\cdot 1024$ is the total number of frequency channels added up from all sidebands and feeds, and $n_\mathrm{samp}$ is the number of samples in time, such that
\begin{equation}
D = 
\begin{bmatrix}
	D_{11} & \dots  & D_{1n_\mathrm{samp}} \\
	\vdots & \ddots & \vdots \\
	D_{n_\mathrm{freq}1} & \dots  & D_{n_\mathrm{freq}n_\mathrm{samp}}
\end{bmatrix}.
\end{equation}
The empirical data covariance matrix, $C$, may then be written as
\begin{equation}
    C = D^TD,
\end{equation}
and the eigenvectors, $\mathbf{v}_k$, of this matrix that correspond to the highest eigenvalues are precisely the PCA components we are looking for. In practice, we identify the few leading PCA components through a standard iterative method. 

For each frequency (in each feed) we compute the PCA amplitudes by projecting the observed data vector, $\mathbf{d}$, onto the PCA eigenvector,
\begin{equation}
	a_k = \mathbf{d}\cdot\mathbf{v}_k  = \sum_{i=1}^{n_\mathrm{samp}} d_i v_k^i,
\end{equation}
where $\mathbf{d}$ is now the normalized, Az/El template subtracted, and polynomial-filtered data described above. The leading PCA components are then subtracted from the data,
\begin{equation}
	\mathbf{d}_\mathrm{after} = \mathbf{d}_\mathrm{before} - \sum_{i=1}^{n_\textrm{comp}} a_k\mathbf{v}_k,
\end{equation}
where $n_\textrm{comp}$ is the number of leading components removed (typically four).

\begin{figure}[t!]
    \includegraphics[width=\linewidth]{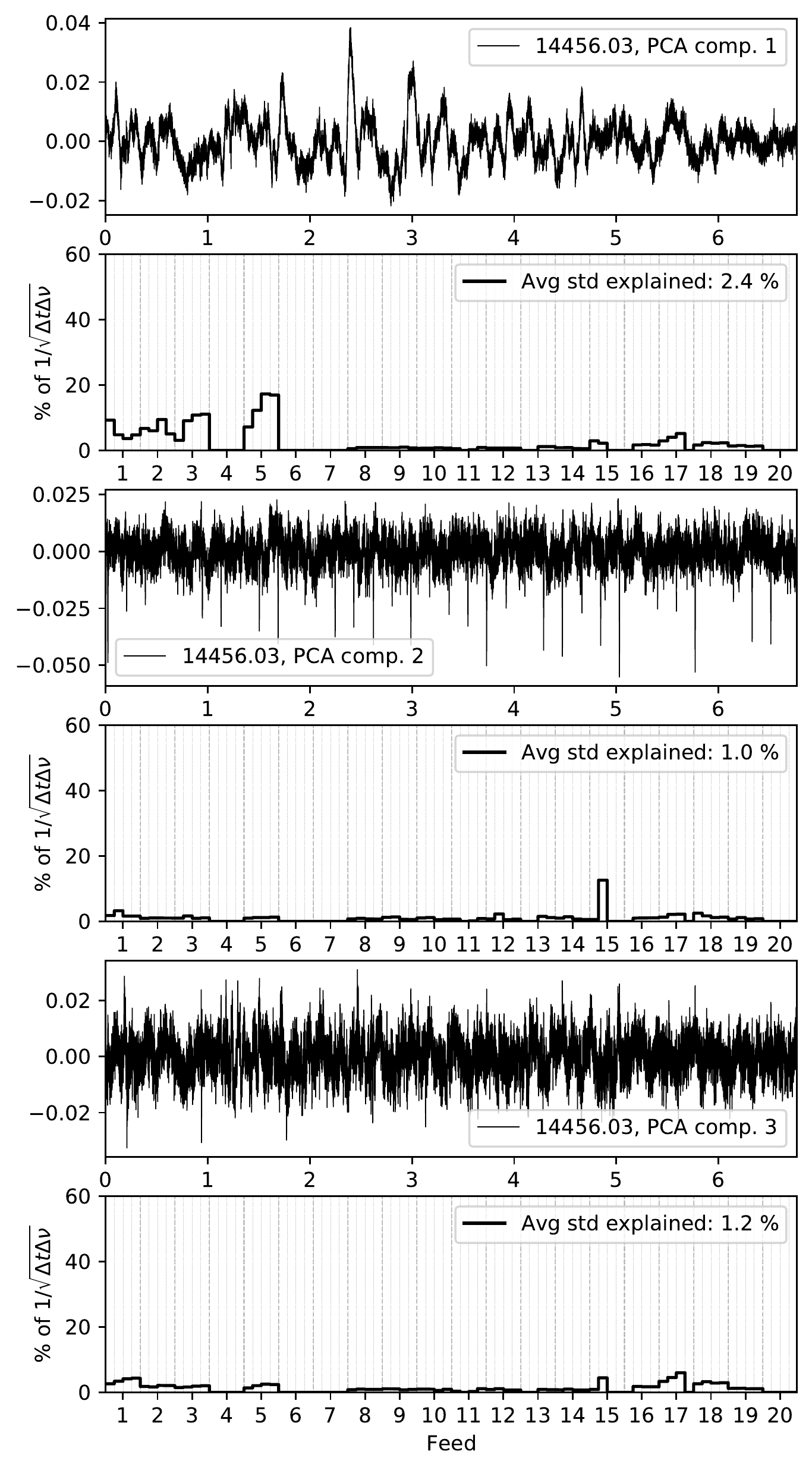}
    \caption{The three leading PCA components of a typical scan, and which feeds are affected.  \label{fig:pca}}
\end{figure}

Figure~\ref{fig:pca} shows the three leading PCA components for a typical scan. For each component, its variation with time is shown for the duration of the scan, as well as its contribution to the overall variance for each feed. Although the contribution of even the leading PCA modes to the overall variance of a typical scan is on the level of single-digit percentages, recall that thermal noise will always dominate the variance for each scan and the spectral structure of even single-digit percentage PCA modes will surely dominate over the targeted CO signal, which is why this filter is important.

\begin{figure*}
    \centering
    \includegraphics[width=\linewidth]{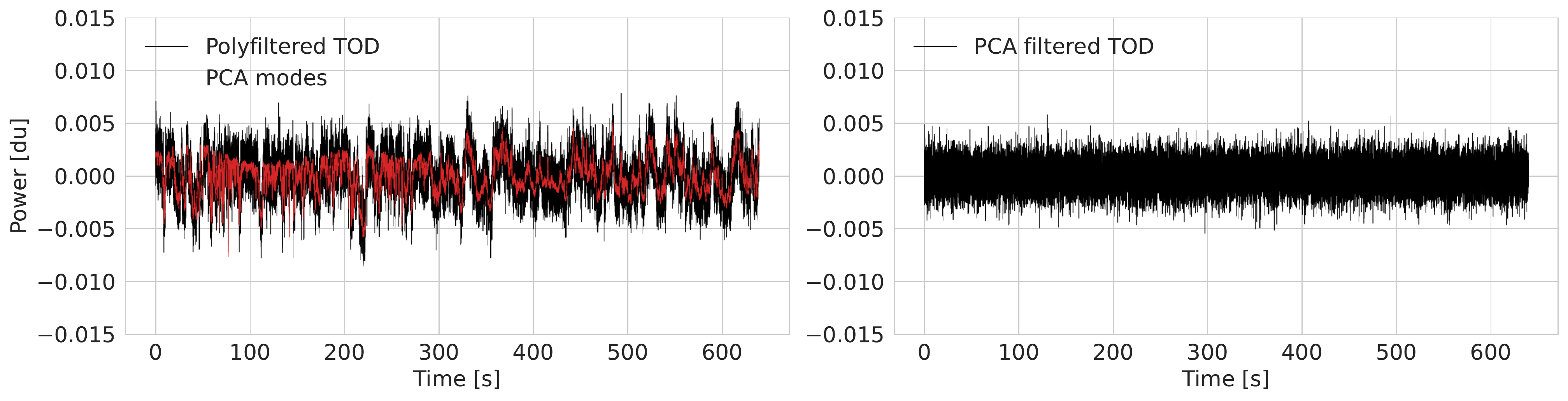}
    \caption{Effect of PCA filter on a ``bad'' scan, with unusually heavy weather or standing wave contributions.}
    \label{fig:TOD_pipeline_PCA}
\end{figure*}

\begin{figure*}[t]
    \centering
    \includegraphics[width=\linewidth]{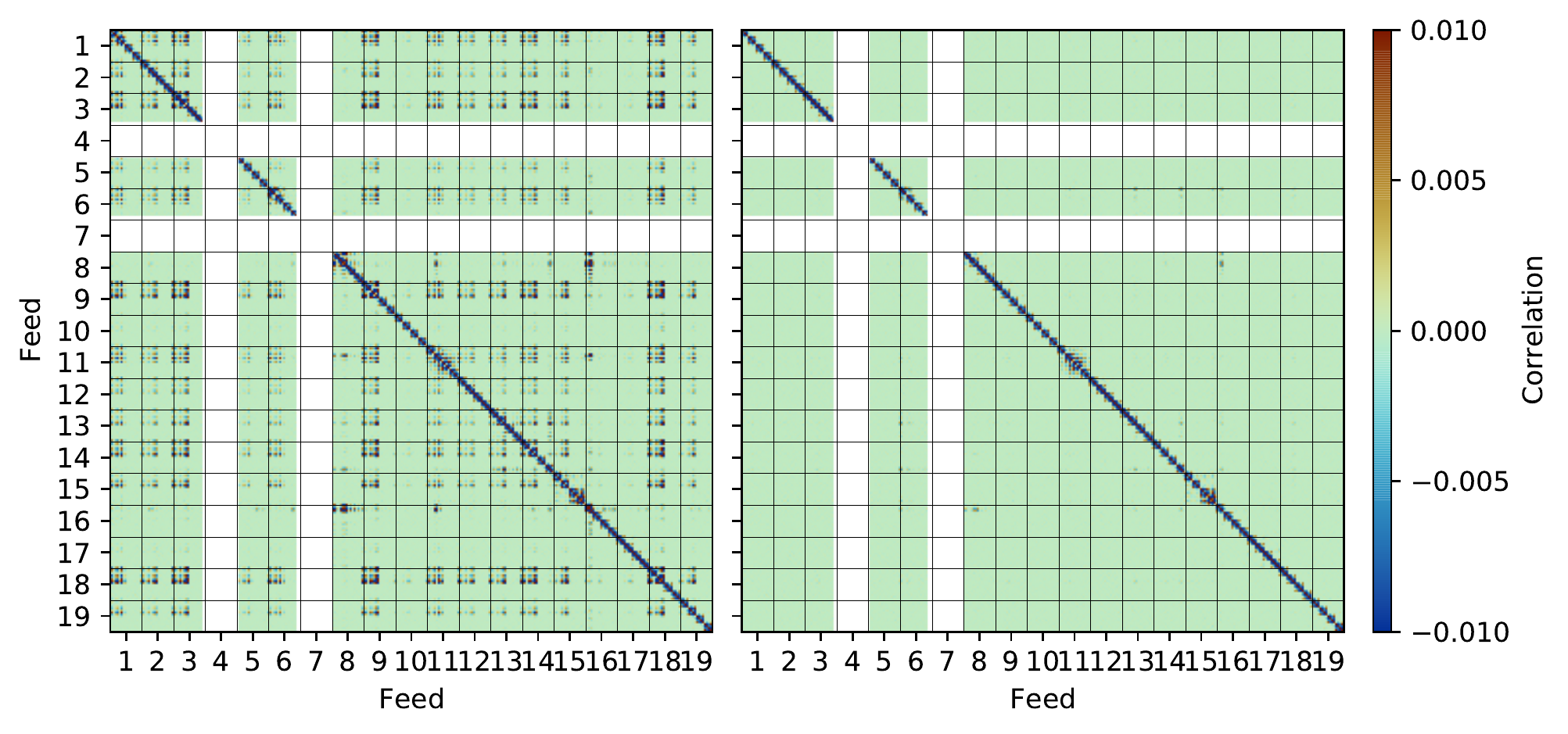}

    \caption{Comparison of channel--channel correlation matrices before (left panel) and after (right panel) applying the PCA filter.}
    \label{fig:corr_filter}
\end{figure*}

Figure~\ref{fig:corr_filter} shows the frequency channel to frequency channel correlation matrix between all frequencies of all feeds before and after applying the PCA filter for a single scan. We see that most of the residual correlations between different feeds are removed in the PCA filter. 
A more extreme example, showing a case where the PCA mode dominates the variance of the data, is shown in Figure~\ref{fig:TOD_pipeline_PCA}.

\subsubsection{Masking}
\label{sec:masking}
Sometimes individual frequency channels or groups of nearby frequency channels show artifacts, even after applying all the filters described above. This could manifest in a significant excess noise that is correlated in time, or in correlations between different frequency channels. We wish to mask these frequency channels so that their contribution does not contaminate the final results. 

To determine which frequencies should be masked we first perform the poly- and PCA filters on a copy of the original dataset. We then use two main approaches to identifying individual or groups of frequency channels to be masked. The first approach uses the fact that the expected correlation between two independent Gaussian variables (for large $n_\mathrm{samp}$) is given by $1 / \sqrt{n_\mathrm{samp}}$, where $n_\mathrm{samp}$ is the number of samples used to calculate the correlation. Thus, after accounting for the expected correlation induced by the polyfilter, we know the statistics describing \emph{good} data, and can identify bad data as deviations from these statistics. Specifically, we consider groups of elements within the frequency-frequency correlation matrix (either squares of different sizes or sets of columns), and compare the average absolute correlation within this group with the scatter expected from white noise alone. Any channel with an absolute correlation larger than $5\sigma$ is removed from further analysis. 

Our second approach is to calculate a set of diagnostics for individual frequency channels, for instance the average correlation of the channel in question to all the others in the same sideband, or the average absolute value of the same. We then compare the values of these diagnostics for the different channels and remove significant outliers. 

In addition to these approaches we also remove frequency channels heavily affected by aliasing. This typically corresponds to about 10 \% of the frequency channels, found at the edges of the bands. We mask all frequencies with a suppression of the aliased signal of less than 15 db. For more details on the aliasing effect see \citep{es_II}. We also mask out individual frequency channels with very high system temperatures, (above 80\,K). 

After the full mask has been determined, we apply the mask to the original (unfiltered) dataset, and repeat the filtering described above, but now only using the unmasked data. This prevents bad data from contaminating good data through the various non-local filters.

\subsection{Calibration}\label{sec:gain}
With cleaned and co-added TOD in hand, the final step we need to perform at the TOD level is calibration; that is, assigning a noise temperature scale to the detector readout. From Equation~(\ref{eq:Tsys}), the overall noise level is proportional to $T_\mathrm{sys}$. 

Ideally, in order to calibrate our instrument, we would put a load of a known temperature in front of the telescope and above the atmosphere and compare the measured output power with the output power measured with no load. A good approximation to this is to use an ambient temperature load that covers the receiver feedhorn. Assuming that the telescope, ground, and the atmosphere have the same physical temperature as the ambient load, the output power will be the same as if the load was above the atmosphere \citep{Penzias+1973}. Taking into account the vertical temperature profile and the distribution of the absorbing components in the atmosphere, the corrections are only a few percent for the relevant wavelengths.

In this approach we define the system temperature, $T_\mathrm{sys}$, by
\begin{align}\label{eq:tsys_phys}
    T_\mathrm{sys} \equiv \frac{e^\tau}{\eta_\mathrm{spill}} \Big[&T_\mathrm{rx} + (1 - \eta_\mathrm{spill})T_\mathrm{gnd}  \nonumber\\
    &+ \,\eta_\mathrm{spill} (1 - e^{-\tau}) T_\mathrm{atm} \nonumber\\ 
    &+ \,\eta_\mathrm{spill} e^{-\tau} T_\mathrm{CMB}\Big],
\end{align}
where $T_\mathrm{rx}$ is the noise temperature of the receiver, $\tau$ is the optical depth of the atmosphere and $1 - \eta_\mathrm{spill}$ is the fraction of the astrophysical signal lost to ground spillover. $T_\mathrm{gnd}$ and $T_\mathrm{atm}$ are the physical temperatures of the ground and the atmosphere, respectively, while $T_\mathrm{CMB}$ is the CMB monopole (we neglect other sky contributions). The overall factor of $e^\tau / \eta_\mathrm{spill}$ converts from a system temperature defined at the receiver input to one defined outside the atmosphere. This definition ensures that 
\begin{equation}
    \Delta T_\mathrm{sys} = \Delta T_\mathrm{signal},
\end{equation}
meaning that a change $\Delta T_\mathrm{signal}$ in the sky signal gives a corresponding change $\Delta T_\mathrm{sys}$ in the system temperature.
This definition makes the interpretation of our measurements easy and intuitive. 

To measure the system temperature we compare the readout when we have an ambient vane $P_\mathrm{amb}$ in front of the receiver and when we look at the cold sky $P_\mathrm{cold}$. From Equations~(\ref{eq:PTsys}) and \ref{eq:tsys_phys} we can estimate $T_\mathrm{sys}$ as
\begin{equation}
	T_\mathrm{sys} = \frac{T_\mathrm{amb} - T_\mathrm{CMB}}{P_\mathrm{amb}/P_\mathrm{cold} - 1},
\end{equation}
where $T_\mathrm{amb}$ is the ambient temperature and $T_\mathrm{CMB}$ is the cold sky temperature. We then multiply the data $d(\nu,t)$ with the $T_\mathrm{sys}$ measurement to go from (normalized) detector units to temperature
\begin{equation}
	d(\nu,t)\bigg|_\mathrm{K} = d(\nu,t) \langle T_\mathrm{sys}(\nu,t) \rangle.
\end{equation}
This method of calibration allows us to account for both the atmospheric absorption and ground spillover, without having to measure $\tau$ and $\eta_\mathrm{spill}$ themselves. 

\begin{figure}
    \centering
    \includegraphics[width=\linewidth]{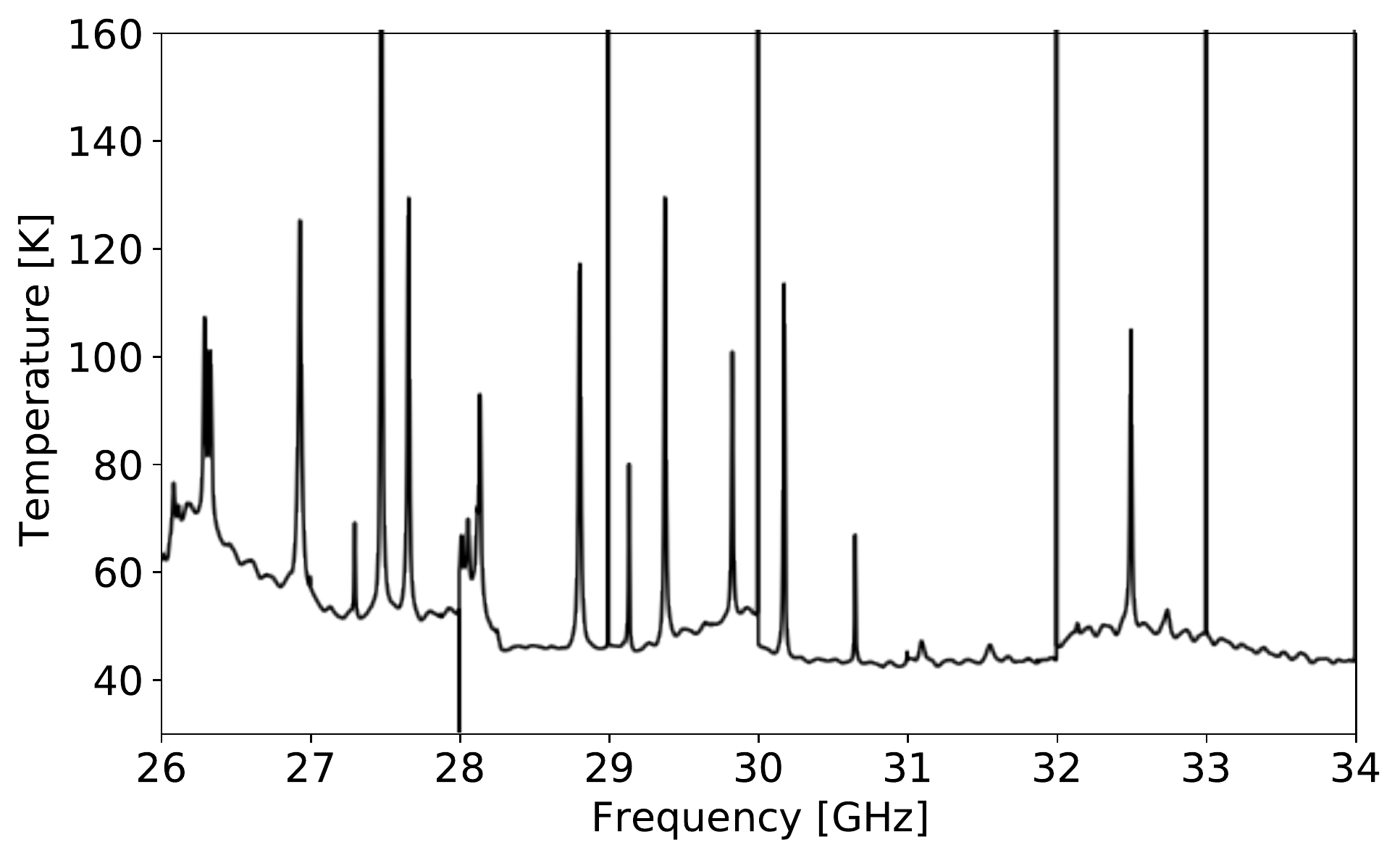}
    \caption{$T_\mathrm{sys}$ measurement from Feed 1 of obsID 15117 across the 4096 frequency channels.}
    \label{fig:tsys_measure}
\end{figure}

As both the atmosphere and the receiver gain vary over time, the measurements of $T_\mathrm{sys}$ vary over time. To get the most accurate estimation, we make use of the ambient vane at the beginning and end of each observation. Figure \ref{fig:tsys_measure} shows a typical example of how the estimated $T_\mathrm{sys}$ looks for a single obsID, as function of frequency. The temperature usually behaves as a relatively smooth function, with large spikes at specific frequencies (for more details see \citealt{es_II}). To see the distribution of sideband averaged system temperature for all Season 1 obeservations we refer you to Figure~\ref{fig:tsys_range} in the Appendix. 

A challenge with this calibration method, is that we are calibrating the total power of the instrument, integrated out to about 90 degrees, rather than just the power in the main beam. As we are interested in structures at small angular scales, some of the total power is essentially lost, with the details depending on the structure of the beam, and the scales of interest. In the power spectrum analysis \citep{es_IV} we take this into account by using a beam transfer function, calibrated on measurements of astrophysical calibration sources \citep{es_VI}.

\subsection{Downsampling}

Until now, all steps have been performed at full frequency resolution, i.e., 1024 channels per sideband or 2\,MHz channel bandwidth. For map-making purposes, however, we typically do not require such high resolution, as the intrinsic line width of the CO signal limits the amount of information at small line-of-sight scales \citep{Chung2021a}. To save both memory and computing time, we therefore co-add several neighboring high resolution frequency channels (usually 16, corresponding to a final bandwidth of 31.25\,MHz) into a single low resolution channel using inverse variance noise weighting. 

\subsection{Map-making}\label{sec:mapmaking}
After the main data selection step (described in Section~\ref{sec:selection}), the last step in the pipeline is map-making, which is implemented in a code called \texttt{tod2comap}. This reads in cleaned TOD and pointing information, applies a highpass filter, and produces temperature sky maps for each frequency channel.
The highpass filter removes structures on long time scales in the TOD. This is done by Fourier transforming the TOD, and removing the part with frequency below a set value, typically 0.02\,Hz, before transforming back to TOD.

Ideally, the TOD can be written as a sum of the signal $\mathbf{s}$ and the noise $\mathbf{n}$,
\begin{equation}
  \mathbf{d}=\mathbf{Ps}+\mathbf{n},
  \label{eq:data_model}
\end{equation}
where $\mathbf{P}$ is the pointing matrix, which connects each time sample to a pixel on the sky. Our goal is to estimate $\mathbf{s}$ given $\mathbf{d}$. Assuming that the noise is Gaussian distributed with a time-domain covariance matrix $\mathbf{N}$, the log-likelihood function corresponding to Equation~(\ref{eq:data_model}) may be written as 
\begin{equation}
	\log\mathcal{L} \propto \left(\mathbf{d}-\mathbf{Ps}\right)^\mathrm{T}\mathbf{N}^{-1}\left(\mathbf{d}-\mathbf{Ps}\right).
\end{equation} 
Setting the derivative of this log-likelihood to 0, we obtain the standard map-maker equation,
\begin{equation}
  \mathbf{\hat{s}} = \left( \mathbf{P}^\mathrm{T}\mathbf{N}^{-1}\mathbf{P} \right)^{-1}\mathbf{P}^\mathrm{T}\mathbf{N}^{-1}\mathbf{d}.
  \label{eq:mapmaking}
\end{equation}
As discussed above, the COMAP noise after filtering is very close to white, and this implies that $\mathbf{N}$ may be approximated as diagonal.\footnote{This is not strictly correct for long time-scales. As such, the current map-maker is statistically slightly sub-optimal, and the resulting transfer function is lower than strictly necessary. Future implementations of the COMAP map-maker will therefore instead rely on well-established destriping or maximum-likelihood algorithms, which are often able to recover slightly more large-scale information than a binning map-maker.} In that case, Equation~(\ref{eq:mapmaking}) may be solved explicitly and independently for each pixel $p$ as follows,
\begin{equation}\label{eq:map}
	\mathbf{\hat{s}}_p = \frac{\sum_{t\in p} \sigma_t^{-2}d_t}{\sum_{t\in p} \sigma_t^{-2}}.
\end{equation}
Here $\sigma_t$ is the noise standard deviation of sample $t$, and samples with lower noise are thus weighted more strongly than the samples with higher noise.
The corresponding map-domain noise standard deviation is given by
\begin{equation}
	\sigma_p = \left(\sum_{t\in p} \frac{1}{\sigma_t^2}\right)^{-1/2}.
        \label{eq:stddev}
\end{equation}
We perform this map-making procedure separately for each frequency channel. 

\section{Data Selection, observation efficiency}\label{sec:selection}
As we will show in Section~\ref{sec:results} most of our filtered timestreams are dominated by white noise. However, this does not necessarily imply that they are free
from systematic errors to a level required for scientific analysis. On the
contrary, many effects may only be discovered when co-adding both over time and frequency.

The main challenge for data selection is to identify and remove data contaminated by systematic errors. It is preferable to remove bad data at the earliest stage possible, before they are co-added with clean data. However, co-adding data also reduces the noise, making it easier to identify systematic effects at a later stage. For this reason, since we cannot detect all systematic errors during the low-level filtering and masking, we go through several stages of data selection, throughout the data analysis pipeline.

In addition to the frequency masking described in Section~\ref{sec:masking}, we also apply cuts based on statistics calculated for each sideband of each feed and scan. These statistics allow us to find patterns and correlations at levels far below the noise level of an individual scan. 

\subsection{Data losses and efficiency}

In order to quantify the overall data efficiency, $E_\mathrm{data}$, i.e. the fraction of raw data we use for the final power spectrum estimates, we summarize the different stages at which data are rejected:

\begin{enumerate}
	\item No data from feeds 4, 6 and 7 were used for the final analysis. Two of these feeds (4 and 7) were used for engineering tests and did not produce useful data, while large systematic errors were visible in the low-level data for one of them (feed 6). We denote the fraction of data lost by rejecting data from these feeds as $L_\mathrm{feed}$. 
	\item As described in Section~\ref{sec:masking}, during the low level data filtering (in \texttt{l2gen}), we mask  bad, outlier or aliased frequency channels. This means that we lose some fraction of the data, denoted by $L_\mathrm{freq}.$
	\item During the first year of observations, we took a large amount of data at elevations above 65 and below 35 degrees. Since we now know that these data are very susceptible to ground contamination (via the main beam and sidelobes), we do not use these data in our results. The fraction of data lost in this elevation cut is denoted by $L_\mathrm{el}$. 
	\item In the main data selection stage (in \texttt{accept\_mod}), we reject data based on a large list of statistics and housekeeping data, including probability of bad weather, large system temperatures, $T_\mathrm{sys}$, pickup of the sun in sidelobes, and many other statistics. We denote the fraction of the data that is lost at this stage by $L_\mathrm{stats}$. 
	\item In \texttt{accept\_mod} we also reject data based on large excesses in the measured power spectra (calculated from individual sidebands, feeds, scans and obsids) (the so-called \texttt{ps\_chi2} statistics). We denote the fraction of data that is lost at this stage by $L_{\chi^2_{P(k)}}$.
	\item During the final stages of the main power spectrum estimation we calculate the cross-spectra for data from different feeds and different data splits (for more details on this see the companion paper \citealp{es_IV}). For each of these spectra we calculate $\chi^2$ statistics that are used to accept or reject the spectra in the final results. This allowed us to identify problems associated with specific feeds. For example, feed 8 had a known problem with the LNA and almost all cross-spectra involving this feed had a high $\chi^2$ statistic; we therefore removed all spectra involving this feed at this step. For Field 1 we also found clear excesses in several spectra involving feeds 16 and 17 from the low elevation dataset; all these spectra were also removed from Field 1. We also cut any spectrum with a $\chi^2$ above 5\,$\sigma$. We denote the fraction of data that is lost at this stage by $L_{\chi^2_{C(k)}}$.
	\item Power spectra formed from the data taken using the Lissajous scanning strategy (which we used for about half of the observations in the first season) showed strong large-scale excess power, potentially due to ground contamination, which is more easily removed from CES scans. For this reason, we did not include any of the Lissajous data in our final science results. We denote the fraction of the final data using CES scans as $E_\mathrm{scan}$. 
	\item Finally, there are periods in time where the telecope, for whatever reason, is not observing the main science CO fields. We denote the fraction of time that we are taking CO data by $E_\mathrm{obs}$. For each of our three individual CO fields we define the observation efficiency as the total time the field was observed multiplied by three and divided by the total period over which the observations were taken. 
\end{enumerate}

\begin{table*}
	\begin{center}
		\caption{Losses and overall data efficiencies for the first season data (columns 1 -- 4) and 
		an optimistic projection for years 2-5 (column 5). Losses, $L$, denote the fraction of data lost at each step. 
		Here $E_\mathrm{data}$ is the product of the factors $(1 - L)$ for all the losses in the rows above.}
		\label{tab:losses}
				\begin{tabular}{l|c|c|c|c||c|c}
			 & Field 1 & Field 2 & Field 3 & All fields & Projection\\
			 \hline
			$L_\mathrm{feed}$ & 15.8\,\% & 15.8\,\% & 15.8\,\% & 15.8\,\% & 0.0\,\% \\
			$L_\mathrm{freq}$ & 26.7\,\% & 28.1\,\% & 26.7\,\% & 27.2\,\% & 15.0\,\% \\
			$L_\mathrm{el}$ & 7.3\,\% & 31.2\,\% & 29.5\,\% & 24.4\,\% & 0.0\,\% \\
			$L_\mathrm{stats}$ & 47.4\,\% & 35.7\,\% & 44.9\,\% & 42.6\,\% & 35.0\,\% \\
			$L_{\chi^2_{P(k)}}$ & 20.9\,\% & 22.3\,\% & 40.0\,\% & 27.8\,\% & 20.0\,\% \\
			$L_{\chi^2_{C(k)}}$\footnote{These are the losses for the CES cross spectra, which are the only ones we ended up with using in the final results. The corresponding losses for the Lissajous data are given in \citet{es_IV}.} & 24.6\,\% & 78.8\,\% & 39.6\,\% & 47.6\,\% & 10.0\,\% \\
			\hline
			$E_\mathrm{data}$ & 18.0\,\% & 4.42\,\% & 8.70\,\% & 9.50\,\% & 39.8\,\% \\
		\end{tabular}
	\end{center}
\end{table*}

\subsection{Future sensitivity projections}

Table~\ref{tab:losses} shows the data lost at different stages of data selection, as well as an optimistic projection for how these values could change in the future. As we can see, a large fraction of the data is lost in the final stage of cuts (based on the $\chi^2_{C(k)}$ statistics), indicating that there are systematic errors that are not being identified in earlier steps. By understanding the origin of these errors and removing them at an earlier stage, there is the potential to significantly increase the amount of data available for analysis.

As mentioned, in the case of data taken using Lissajous scans (which corresponds to about half of the total obtained in Season 1), there is a clear excess in the final power spectrum; for this reason, these data were not used for our science results. 
For data taken using CES scans, the $\chi^2_{C(k)}$ cut produces spectra for Fields 1 and 2 that are consistent with white noise. For Field 3 we needed to apply a more restrictive set of limits on the various statistics and $\chi^2_{P(k)}$, which we believe to be related to an increased level of ground contamination compared to the other two fields.

With experience of Season 1 in hand, we are working on building the second generation COMAP pipeline, including improved ground modeling, map-making and real time continuum filtering and calibration, based on the lessons learned from the first generation data analysis and our improved understanding of the data. We have also altered our observing strategy and corrected hardware problems and the combined effect of these improvements is expected to lead to a significant increase in sensitivity. We discuss the expected improvement in each factor from Table~\ref{tab:losses} below.

\begin{enumerate}
	\item At the end of Season 1, the receiver was removed from the telescope for maintenance. Feeds 4 and 7 were switched from engineering testing to science operations mode, while problems with feed 6 were repaired. In subsequent observing seasons, we therefore hope to keep $L_\mathrm{feed}$ close to 0\,\%. 
	\item We plan to increase the clock frequency of the analog-to-digital converters in the ROACH-2 backend. This will reduce the number of frequency channels removed due to aliasing and, coupled with improvements in the filtering and a more stable system, should improve $L_\mathrm{freq}$ significantly. 
	\item Regarding $L_\mathrm{el}$, we are no longer observing above 65 and below 35 degrees in elevation, so we should not lose any more data to this elevation cut in subsequent seasons. 
	\item With a careful study of the effect of relaxing the current conservative cuts based on the various statistics, we believe that there is some scope to reduce $L_\mathrm{stats}$. 
	\item We believe it will be possible to significantly improve $L_{\chi^2_{P(k)}}$ if we choose the limits more carefully.
	\item By identifying data affected by systematic errors at an earlier stage in the pipeline and by splitting the data into more pieces for the cross-correlation, we expect to be able to significantly reduce $L_{\chi^2_{C(k)}}$ (especially for Fields 2 and 3). Field 1 shows that it is possible to reduce the losses at this step significantly. For Field 1, after we removed all spectra involving feed 8 and all spectra involving feeds 16 and 17 from the low-elevation dataset by hand, the automatic $\chi^2_{C(k)}$ cut at 5$\sigma$ accepts all but one of the remaining 182 cross spectra, indicating that the remaining data are very clean.
	\item After Season 1 we no longer use Lissajous-type scans, which means that $E_\mathrm{scan} = 100 \%$ for all subsequent seasons. 
	\item During Season 1, we addressed the main instrumental and operational issues that decreased $E_\mathrm{obs}$, as well as instituting weekly maintenance checks; we expect to achieve close to the maximum efficiency of 82.5 \% (based on the total time our CO fields are within 35--65 degrees elevation) for future seasons. We will therefore assume an observation efficiency of 75 \% for the future, a large improvement over the values, 36.8\,\% 52.9\,\% and 53.2\,\% for Fields 1, 2 and 3, respectively, obtained during the first season. 
\end{enumerate}

Based on these considerations we make an estimate of our future data efficiency, as shown in Table~\ref{tab:losses} (the rightmost column). The future (five year) power spectrum sensitivity is then estimated simply by scaling the current sensitivity by the appropriate factor, $D$,
\begin{equation}
    \sigma^\mathrm{5yr} = \frac{\sigma^\mathrm{S1}}{D^\mathrm{5yr}}.
\end{equation}

We define the total efficiency for Season 1 as $E^\mathrm{S1}_\mathrm{tot} \equiv E^{S1}_\mathrm{scan} E^{S1}_\mathrm{data}E^{S1}_\mathrm{obs}$, where the ${}^{\mathrm{S1}}$ denotes the quantity from the first season. The fraction of data using the CES scan during the first season, $E^{S1}_\mathrm{scan}$, was given by 51.7\,\% 55.6\,\% and 34.3\,\% for Fields 1, 2 and 3, respectively. 
We assume for the forecast that by splitting the data into more parts (than the two we are currently using), we can improve the sensitivity of the cross spectrum estimator (see \citealp{es_IV} for more details) by a factor of $E_\mathrm{split} = 1.3$ (the asymptotic limit as $N_\mathrm{split} \rightarrow \infty$ is $\sqrt{2}$). 

Using the total duration of the season 1 observing campaign, $T^{S1} = 440$ days,
we base our forecast on the performance of our best field (Field 1), and find the factor, $D^\mathrm{5yr}$, needed to estimate the five year sensitivity, 
assuming that we can make all three of our fields perform as well as Field 1. We also assume that we can improve the transfer function by at least 10 \% on average by improved filtering and map-making, giving us an extra overall factor $E_\mathrm{TF} = 1.1$ We then get
\begin{equation}
    D^\mathrm{5yr} \equiv \frac{T^{S1}E^\mathrm{S1}_\mathrm{tot}/\sqrt{3} + (5\cdot 365 - T^{S1}) E_\mathrm{tot}}{T^{S1}E^\mathrm{S1}_\mathrm{tot}/\sqrt{3}} E_\mathrm{split}E_\mathrm{TF},
\end{equation}
where $E_\mathrm{tot} \equiv E^\mathrm{proj}_\mathrm{data}E^\mathrm{proj}_\mathrm{obs} = 29.8 \%$, and where the $\sqrt{3}$ comes from the fact that we are extrapolating the current sensitivity of Field 1 to all three fields. Here the superscript ${}^\mathrm{proj}$ denote the previously discussed future projections. Inserting the values we find $D^\mathrm{5yr} = 69.4$.
Discussion of the current upper limit and the five year power spectrum sensitivity forecast can be found in the companion paper \citet{es_V}.

\section{Signal loss and the pipeline transfer function}\label{sec:transfer}

The main summary statistic we use to estimate the CO signal is the power spectrum
\begin{equation}\label{eq:PS}
P(\vec{k}) = \frac{\left< |f_{\vec{k}}|^2\right>}{n_xn_yn_z} V_{\mathrm{vox}}.
\end{equation}
This is extracted from the temperature sky maps by, first, computing the 3D Fourier transform of the maps; binning the squared Fourier coefficients according to the wave number, $\vec{k}$; and averaging over all contributions to a given $\vec{k}$-bin. Finally, they are multiplied by the co-moving voxel volume, $V_{\mathrm{vox}}$, and divided by the total number of voxels, $n_xn_yn_z$. Note that each voxel is inverse variance weighted by $\sigma_{p}^{-2}$ as given by Equation~(\ref{eq:stddev}) before computing the Fourier transform, meaning that we are calculating the pseudo-spectrum rather than a regular auto spectrum (we will not mention this distinction in the rest of this paper, but all power spectra discussed here are pseudo-spectra). For more details on the power spectrum methods see \citet{es_IV}.

As the raw data passes through our filtering and map-making procedures, some of the signal is typically lost at each stage, and the maps described in Section~\ref{sec:mapmaking} are therefore biased. In order to estimate and correct for this bias at each scale, $\vec{k}$, we need to estimate the so-called pipeline transfer function, which is simply defined as the power spectrum ratio between the recovered and original signal.

We can estimate this transfer function by adding a signal-only simulation to a pure noise TOD, and then comparing the combined signal-plus-noise simulation output to the true signal-only input. We adopt the raw COMAP TOD as a model for the noise, which in power units are denoted $P_{\mathrm{N}}$. The signal-only contribution is produced by scanning a pre-computed 3D simulation of brightness temperature (using the fiducial model in \citealp{es_V}), $T_\mathrm{sim}(p)$, with the telescope pointing, and we denote this $P_{\mathrm{S}}$. We then add these together in power units, 
\begin{equation}
    P_\mathrm{S+N} = P_\mathrm{N} + P_\mathrm{S} = k_BG\Delta \nu T_\mathrm{sys}\left(1 + \frac{T_\mathrm{sim}}{T_\mathrm{sys}}\right).
\end{equation}
We then separately generate 3D voxel maps from $P_\mathrm{S}$, $P_{\mathrm{N}}$ and $P_{\mathrm{S+N}}$, and from these we compute corresponding 3D power spectrum $P_{\mathrm{S}}(\vec{k})$, $P_\mathrm{N}(\vec{k})$, and $P_{\mathrm{S+N}}(\vec{k})$, following the above procedure. Based on these three spectra, we can finally estimate a scale-dependent transfer function $T(\vec{k})$ as
\begin{equation}\label{eq:transfer}
    T(\vec{k}) = \frac{P_\mathrm{S+N}(\vec{k}) - P_\mathrm{N}(\vec{k})}{P_\mathrm{S}(\vec{k})}.
\end{equation}

Noting that the pipeline filters have very different impact in the angular and frequency directions, it is useful to decompose $\vec{k}$ into parallel (line-of-sight) modes, $k_\parallel \equiv |k_z|$, and the perpendicular (angular) modes, $k_\bot \equiv \sqrt{k^2_x + k^2_y}$. This is the version of the transfer function we use for the main science analysis. However, for simplicity of visualization we will here show several results for the 1D (spherically averaged) version of the transfer function, in addition to the 2D (cylindrically averaged) one. 
\begin{figure}
    \includegraphics[width=\linewidth]{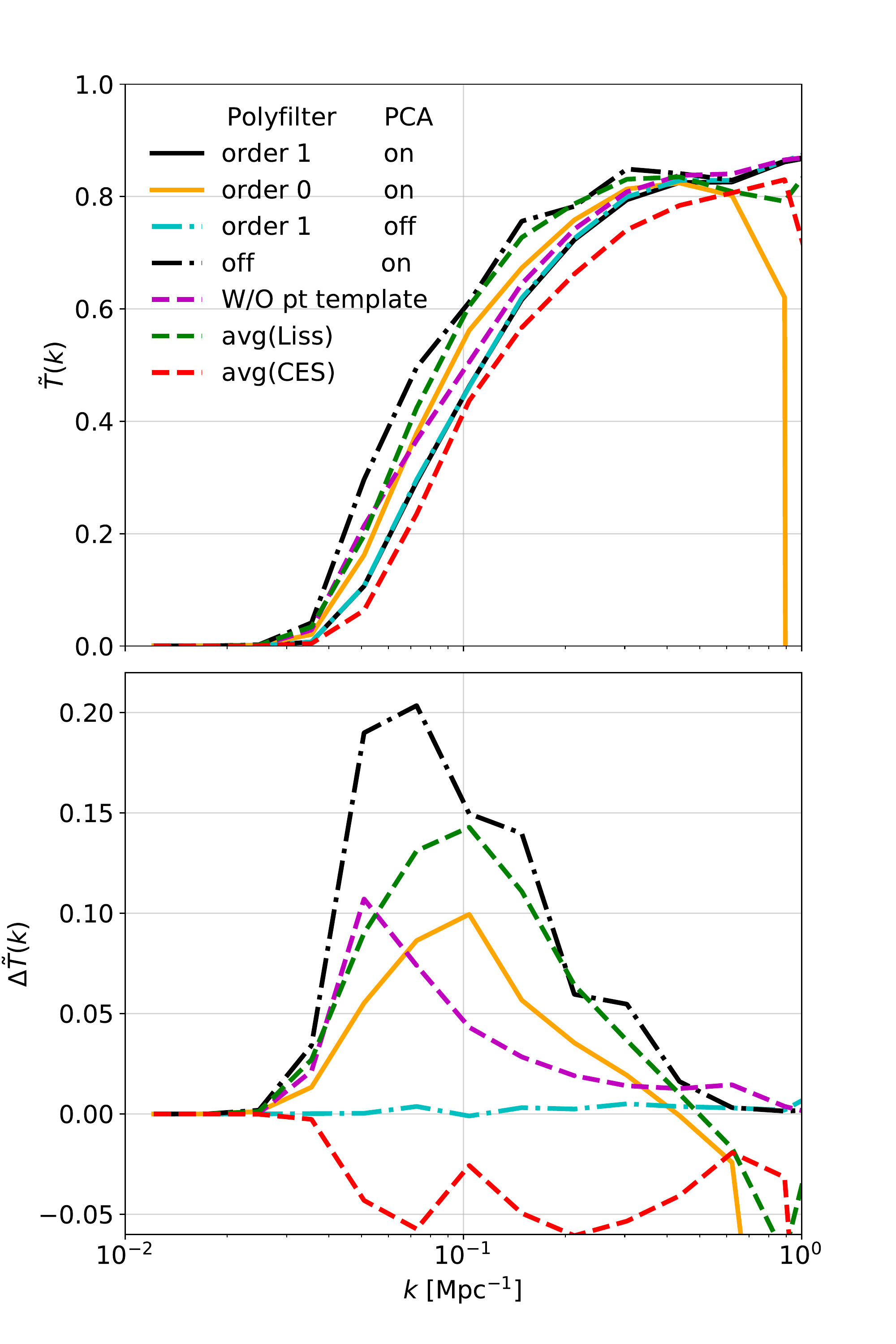}
    \caption{(\emph{Upper panel}:) 1D transfer functions, $T(k)$, for different filter options and scanning modes as a function of scale, $k$. The default combination used in the COMAP pipeline is shown as a solid black line. (\emph{Lower panel}:) Difference between the various filter and scanning options and the default configuration.
    \label{fig:TF_all}}
\end{figure}
\begin{figure*}
    \includegraphics[width=\linewidth]{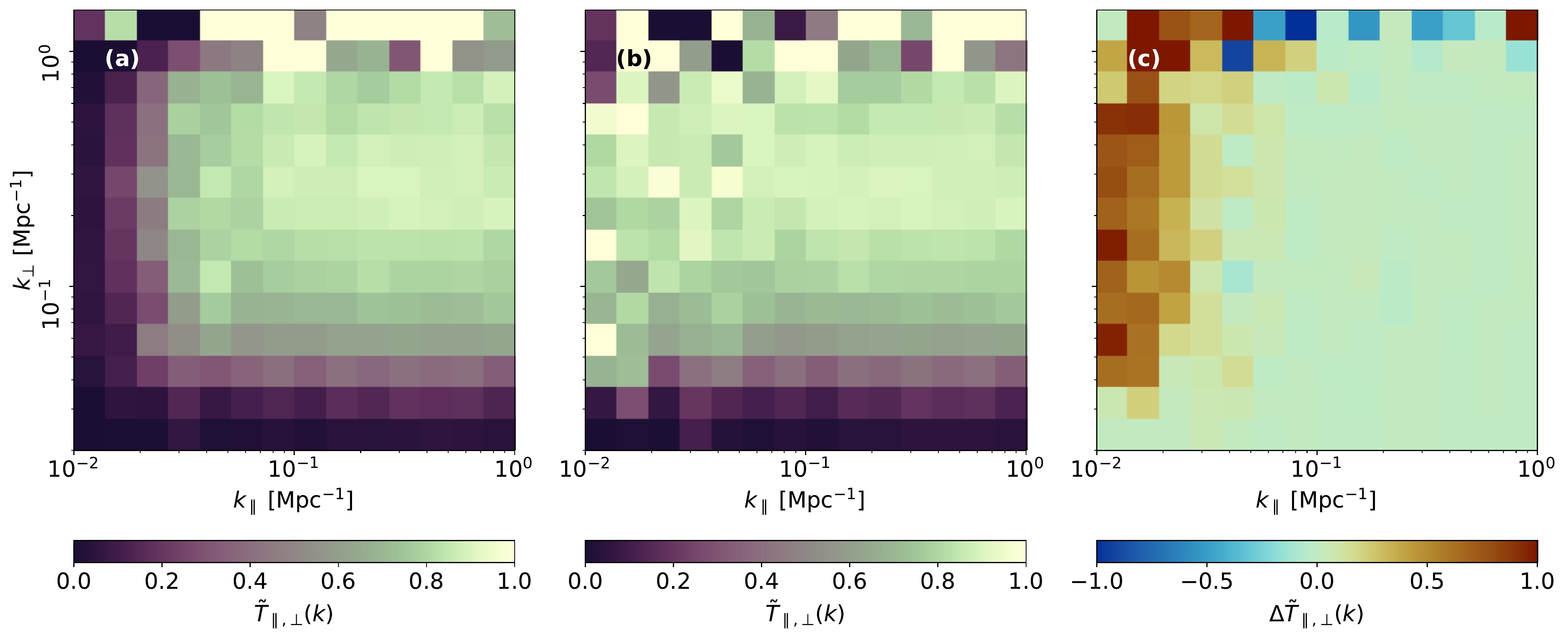}
    \caption{Comparison of 2D transfer function estimates with (\emph{left panel}) and without (\emph{middle panel}) the polyfilter. (\emph{Right panel}:) Difference between the two previous cases.
      \label{fig:TF_default_wopoly}}
\end{figure*}

First, to understand the impact of the various filters in terms of signal-loss, we estimate 1D transfer functions for a range of different pipeline configurations. Specifically, we analyze six obsIDs (three CES and three Lissajous obsIDs), where we consider different combinations of PCA and polyfilter, enabling or disabling each filter in turn. For the polyfilter, we additionally consider two cases, namely a constant fitting term or a linear fitting function. The results from these calculations are summarized in Figure~\ref{fig:TF_all}. The black solid line shows the default pipeline configuration.

One can see that the default settings, i.e., a first order polyfilter and PCA filtering turned on, yields almost the same transfer function as the case where the PCA filter is turned off. The PCA filter is not expected to remove much of the actual input signal, as it only removes the components of the TOD that are the most correlated over all frequencies and feeds, thus potentially removing only the structures of the input signal that are common over the entire survey volume observed at any given time. 

When it comes to the case with a zeroth order polyfilter or with the polyfilter turned completely off there are, however, large differences seen from the results using the default settings. Using a zeroth or first order polyfilter, a considerable fraction of the input signal is removed by the pipeline on scales above $k\sim 0.04\,\mathrm{Mpc}^{-1}$. We see that the zeroth ordered polyfilter yields a similar result as without the polyfilter near the peak regions of the transfer functions, however a non-negligible portion of signal between $k\sim 0.04\,\mathrm{Mpc}^{-1}$ up to the peak region is taken out when turning on the zeroth order polyfiltering. The low transfer efficiency on low $k$ for any of the shown filter combinations is due to the limited area covered in each scan, the highpass filter imposed in the map-maker, as well as the polynomial filter in frequency. 

If we turn off the subtraction of the Az/El templates, we can also see in Figure~\ref{fig:TF_all} that more signal is let through the pipeline on scales $k \leq 0.3\,\mathrm{Mpc}^{-1}$. The effect of the Az/El template subtraction is however especially noticable on scales $k< 0.1\,\mathrm{Mpc}^{-1}$, which is expected as the structures in the power spectra induced removed by the Az/El templates are of a larger scale in the pixel domain.

Note also that when computing these transfer functions for different filter combinations, we used the combined maps of three obsIDs of type Lissajous scan and three with constant elevation scans. However, we found that there were significant differences between the transfer functions from a Lissajous and CES scan type and have therefore also included the average of the three transfer function of each type in Figure~\ref{fig:TF_all}. As one can see, the Lissajous scan type results in a transfer function that is larger on most scales, which probably is a result of the Lissajous scan covering a larger area in a single scan compared to the constant elevation scans. The Lissajous scans, as opposed to the ones with CES, also seem to result in a transfer function that drops a bit down from its peak at high $k$. The reason for this difference is not yet fully understood at this point.

In general the estimates of the transfer function break down at high $k_\bot$ due to $P_\mathrm{S}(\vec{k})$ going to zero in Equation~(\ref{eq:transfer}), so we see some large random fluctuations here, but this is not a problem because the instrumental beam means that we have basically no sensitivity to these modes anyway. 

When looking at the 2D version of the transfer function, as shown in Figure~\ref{fig:TF_default_wopoly}, the effects of the polyfilter on the transfer function become more evident because we can then distinguish between what is happening in the angular directions ($k_\bot$) and the spectral dimension ($k_\parallel$). As the polyfilter is designed to remove the $1/f$-noise as well as continuum foreground emission along the frequency dimension on each sideband, we expect the changes in the transfer function to be most visible in the large line-of-sight scales $k_\parallel$. This is indeed what is seen in the difference $\Delta T(k)$ between the transfer functions without and with a first order polyfilter in Figure~\ref{fig:TF_default_wopoly} for low $k_\parallel$, where we note a $50-90\,\%$ relative loss in power when using a first order polyfilter. Meanwhile on all other scales the difference $\Delta T$ is left mostly unchanged. 

\section{Noise characterization and removal of correlated noise}\label{sec:results}
In this section we describe our noise characterization model, and demonstrate how effectively the pipeline removes correlated noise from the data.
One of the most important functions of the COMAP data pipeline is the removal of correlated noise. By correlated noise, we mean that noise that is correlated in time. The fact that all known sources of correlated noise in our system also produce noise that is correlated across feeds or frequencies, gives us powerful leverage to remove correlated noise from our data. As the CO brightness temperature is many orders of magnitudes below the telescope system temperature, any significant deviations from a white noise spectrum in our filtered data, must be due to residual correlated noise or another systematic effect. This means that in order to quantify the presence of correlated noise, we can look at the deviation from white noise. 

\begin{figure}
    \centering
    \includegraphics[width=\linewidth]{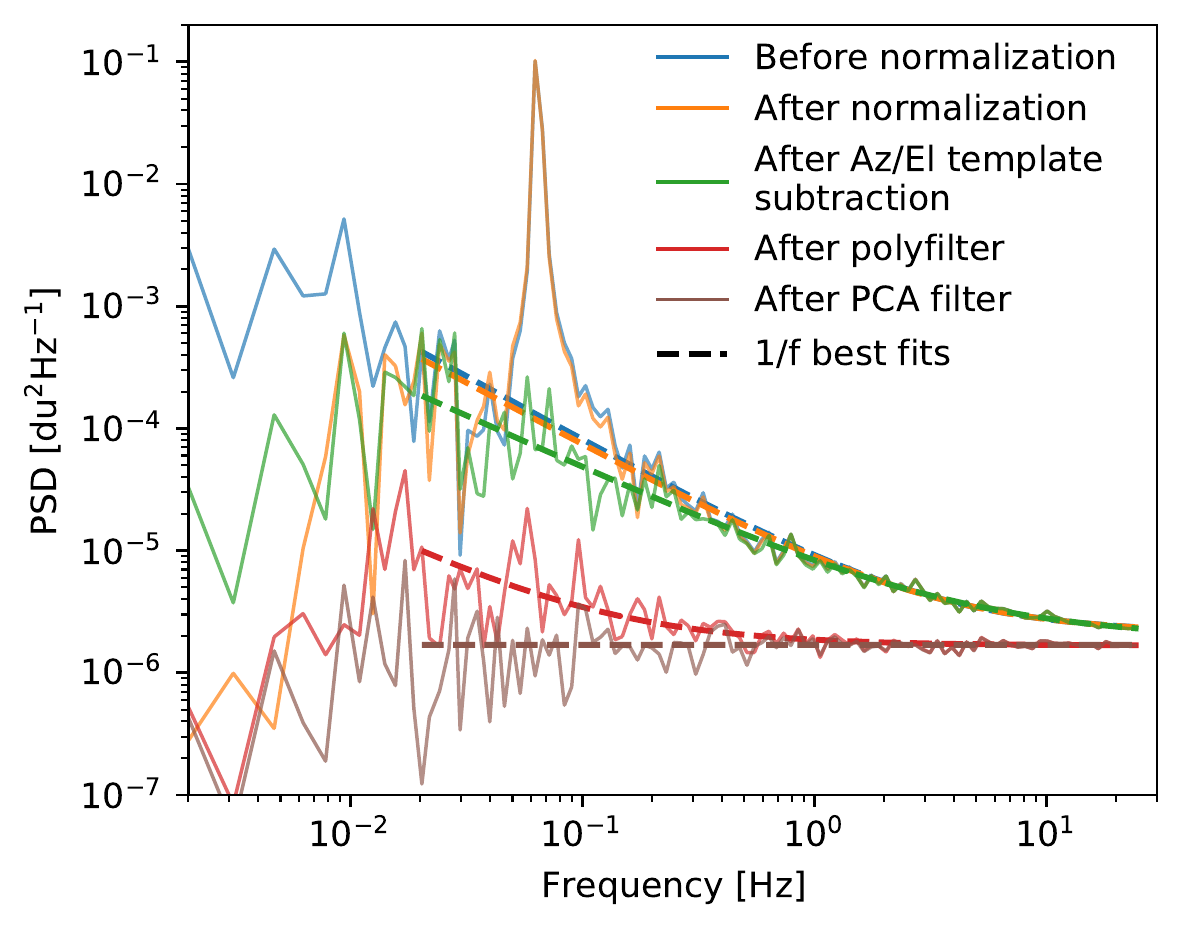}
    \caption{Power spectral distribution of a single scan from a 31.25\,MHz band around 28.2\,GHz at different stages in the pipeline, with $1/f$ noise curves fitted. The power spectral distribution is binned with logarithmic bin-sizes towards higher frequencies for clarity. Lower frequencies have been excluded from the fit, as these scales are greatly suppressed at the normalization stage.}
    \label{fig:1f_fits}
\end{figure}

\begin{figure}
    \centering
    \includegraphics[width=\linewidth]{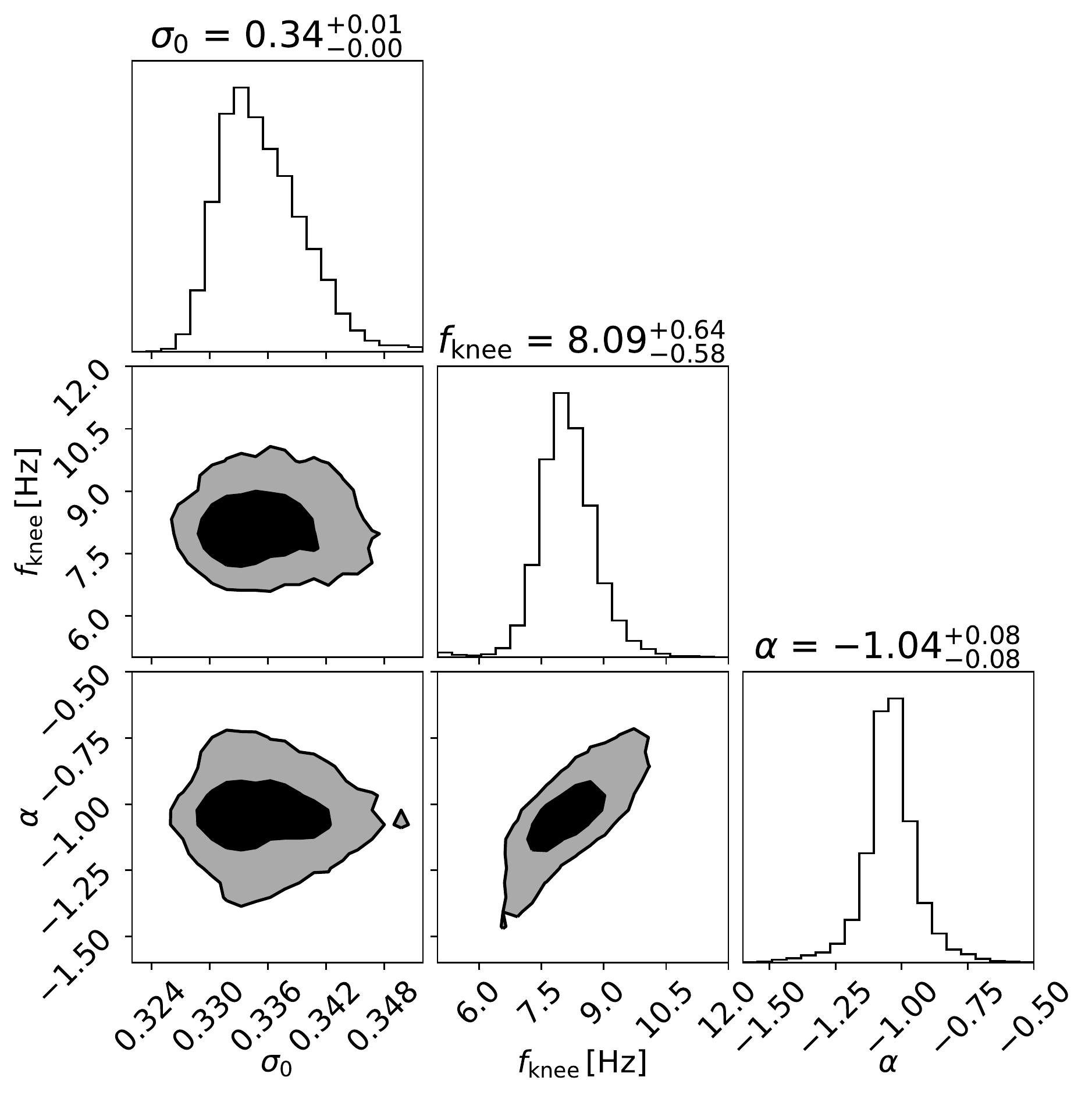}
    \caption{Distribution of noise parameters $\sigma_0$, $f_\mathrm{knee}$ and $\alpha$ for the $c_0$ coefficient of the polyfilter. All available scans of feed 1, sideband A:LSB was used.}
    \label{fig:poly_c0_corner}
\end{figure}

We can often approximate the correlated noise using a spectral density on the form
\begin{equation}\label{eqn:1f_noise}
	N(f) = \sigma_0^2\left(1 + \left(\frac{f}{f_\mathrm{knee}} \right)^\alpha\right),
\end{equation}
where $\sigma_0$ is the white noise level.

The second term in Equation~(\ref{eqn:1f_noise}) is known as $1/f$ noise,\footnote{Keep in mind that $f$ refers to the temporal frequency of the time ordered signal, not the observed photon frequencies, to which we consistently refer to as $\nu$.} which is characterized by a knee-frequency $f_\mathrm{knee}$, representing the transition frequency between the flat white noise and the sloped $1/f$ noise, and the exponent $\alpha$, giving the slope of the spectral density in the $1/f$ dominated regime.
The white noise level is estimated by calculating the variance between neighboring samples in the TOD, as
\begin{equation}\label{eqn:sigma_0}
    \sigma_0 = \sqrt{\frac{\mathrm{Var}(d_i - d_{i-1})}{2}}.
\end{equation}

Figure~\ref{fig:1f_fits} shows the $1/f$ behavior of the TOD throughout different steps in the pipeline, and clearly demonstrates the effect of each filter. The normalization step heavily suppresses the low-frequency end of the spectrum. The Az/El template knocks out the strong $\sim\!0.7$\,Hz correlation caused by the Lissajous scanning strategy. The polyfilter significantly reduces the noise power across the entire power spectrum, even lowering the white noise limit. This is possible because even though the white noise is uncorrelated in time, parts of it are still correlated in frequency. Finally, the PCA filter further reduces the noise left over by the polyfilter. By the end of the pipeline, the TOD is almost completely dominated by white noise. It should be noted that while the polyfilter typically suppresses much more noise power than the PCA filter in an average scan, this is not always the case. In scans with significant contamination (like standing waves or bad weather), the PCA filter may suppress even more noise power than the polyfilter. An extreme example of this is shown in Figure~\ref{fig:TOD_pipeline_PCA}.

\subsection{Polyfilter noise properties}
As discussed in Section~\ref{sec:poly}, the polyfilter involves fitting and subtracting a low order polynomial in frequency space from each sideband at each individual timestep. The polyfilter is the first filter targeting correlated noise except on the very larges timescales, and the resulting coefficients are therefore highly informative regarding the noise properties of the data. In the current analysis setup, we only use a first order polynomial filter, such that each timestep of each sideband are associated with two coefficients, $c_0$ and $c_1$. These coefficients, treated as functions of time, turn out to have $1/f$-like power spectra. Figure~\ref{fig:poly_c0_corner} shows the distribution of noise parameters of $1/f$ fits performed on $c_0$ for all available scans of the A:LSB sideband of feed 1. As discussed in Section~\ref{sec:model} the correlated noise common to each sideband is mostly dominated by gain fluctiations of the individual low noise amplifiers at each feed. We therefore expect, and find, that each feed has its own characteristic noise properties. Since we can use the polyfilter to remove this correlated noise, the individual noise properties of the different feeds are less important when measuring the CO line emission than if we were measuring continuum sources, in which case these properties would become crucial.

\begin{figure}
    \centering
    \includegraphics[width=\linewidth]{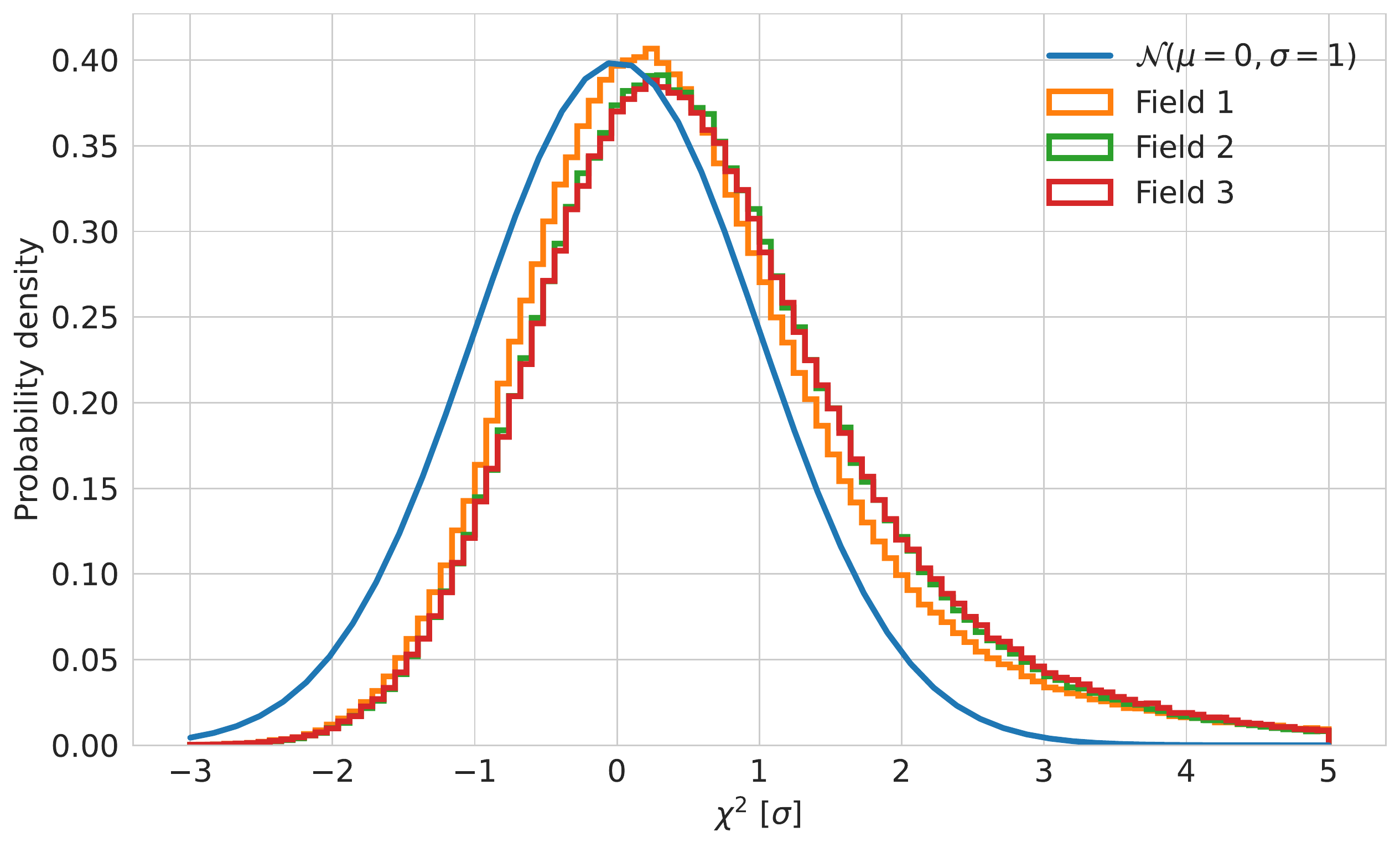}
    \caption{$\chi^2$ distributions of filtered data for the three main fields, with a standard normal distribution for comparison.}
    \label{fig:chi2_hist}
\end{figure}

\subsection{Goodness-of-fit, $\chi^2$-test}
The main goal of our pipeline is to remove both correlated noise and continuum foregrounds, while leaving as much as possible of the CO line intensity signal intact. In the ideal case, and assuming that the cosmological CO signal is so weak that it can not be measured in a single scan, our cleaned TOD should therefore be described by white noise alone. We therefore need statistics to measure potential deviations from white noise. We use a standard $\chi^2$ statistic per scan for this purpose, defined as follows,
\begin{equation}\label{eq:chi2}
    \chi^2 = \frac{\sum\limits_{i=0}^N\left(\dfrac{d_i}{\sigma_0}\right)^2 - N}{\sqrt{2N}}.
\end{equation}
Here $d_i$ are the $N$ samples of the scan, and $\sigma_0$ is the white noise level defined in Equation~(\ref{eqn:sigma_0}). For a perfect white noise TOD, we expect $\chi^2 \sim \mathcal{N}(\mu = 0,\, \sigma=1)$.

Figure~\ref{fig:chi2_hist} shows the $\chi^2$ distribution for all the scans in the first observation season, comprising about 5\,000 hours of observations, divided by observational field. Here we have combined all the datapoints for each sideband, such that the $N=n_\mathrm{samp} \cdot n_\mathrm{freq}$, where $n_\mathrm{samp}$ is the number of samples in time (typically $n_{\mathrm{samp}}=10$--20\,000) and $n_\mathrm{freq}=64$ is the number of frequencies per sideband. As seen in Figure~\ref{fig:chi2_hist}, the data are indeed very close to white noise, with only a small shift and a positive tail. We also note that the Field 1 field outperforms the two other fields by a small margin. Given that the number of samples, $N\sim 10^5$, going into each of the $\chi^2$ values in this histogram is so large, a mean bias of less than 1\,$\sigma$ per scan suggests that the typical residual variance from correlated noise is very low.

\subsection{Maps}
\begin{figure*}
    \centering
    \includegraphics[width=1.0\linewidth]{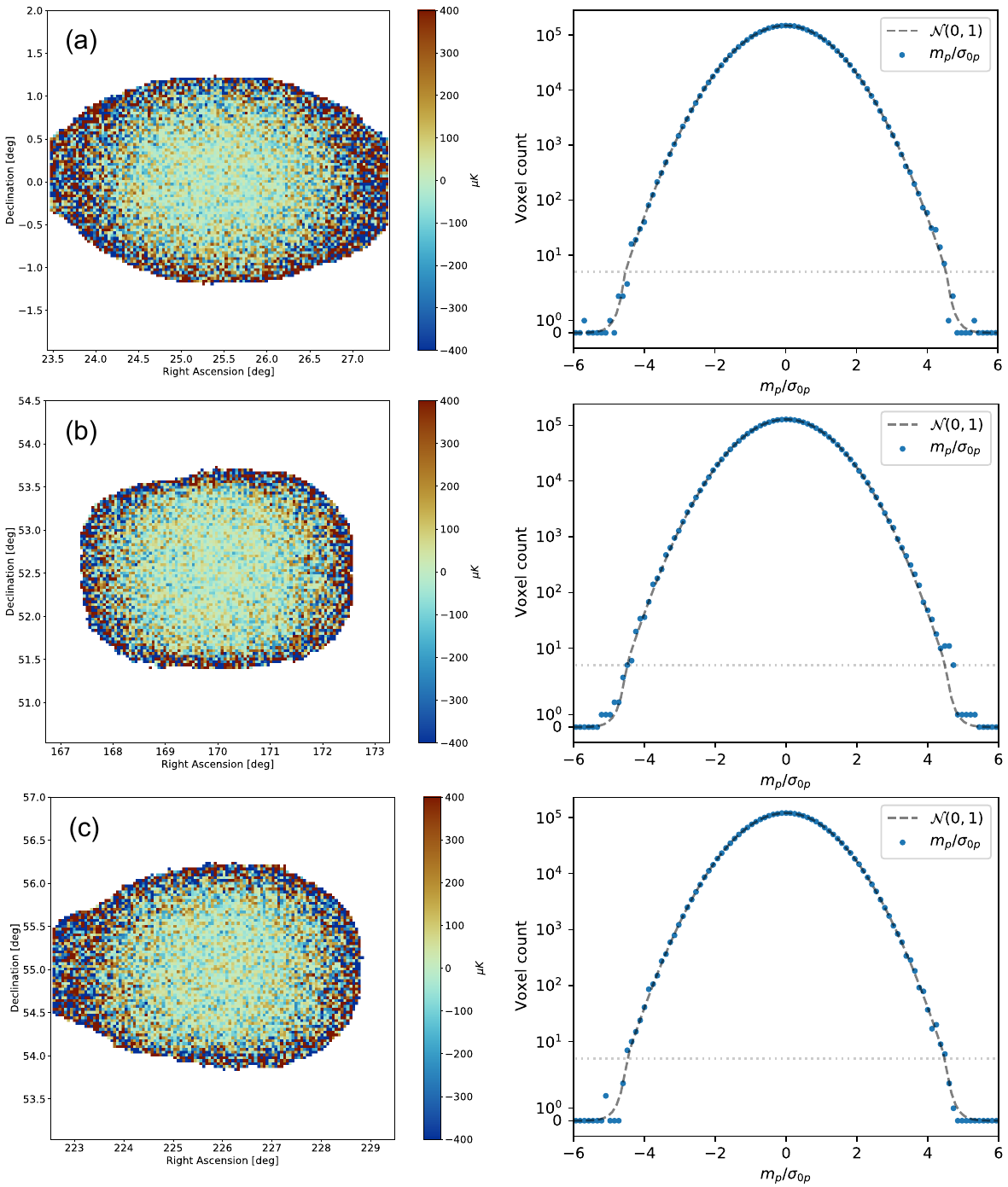}
    \caption{Co-added COMAP single 31.25 MHz frequency channel maps with a central frequency of 28.9 GHz (left) and voxel histograms of the map voxels divided by their corresponding white noise level for all 3D voxels (right) for Field 1 (a), Field 2 (b) and Field 3 (c). Regions that are either not observed by the telescope, or have a noise level $\sigma_\mathrm{0}> 1000\,\mathrm{\mu K}$, are masked out in the plotted maps. Note that in the voxel histogram we use a linear $y$-axis below a voxel count of 5, and a logarithmic one above. \label{fig:maps}}
\end{figure*}

Figure~\ref{fig:maps} shows a single frequency map, from each of our three fields, based on the data from the first season of observations. This data set results in a sensitivity of a few tens of $\mu$K per 2$\times$2 arcmin${}^2$ pixel for a single 31.25\,MHz channel. At least at a visual level, the maps appear largely dominated by white noise. We see that each field only has significant coverage within rougly a 2$\times$2 deg${}^2$ area on the sky. The right panel of the figure shows histograms of all the map voxels, $m_p$ (Equation~\ref{eq:map}), divided by their corresponding white noise level, $\sigma_{0p}$ (Equation~\ref{eq:stddev}). Overplotted is what is expected from a unit normal variable. We see that the individual voxels follow very closely a Gaussian distribution with a standard deviation consistent with the white noise level of the timestreams that make up the map, further illustrating that the maps are dominated by white noise at the single voxel level.

\section{Summary and conclusions}\label{sec:conclusion}

We have presented the data analysis pipeline used to process the first-season COMAP observations with respect to high-redshift CO emission, from raw time-ordered data to final calibrated maps. This pipeline implements four main steps (calibration, filtering, data selection, and map-making), each of which is designed to optimally exploit the unique instrumental capabilities of the COMAP instrument. For instance, calibration is performed using a combination of frequent comparison with a hardware calibrator and real-time total power measurements. The filtering procedures explicitly exploit the multi-feed and -frequency design of the COMAP instrument to reject common-mode contaminants, resulting in data that are strongly dominated by uncorrelated white noise after filtering. Finally, both the data selection and map-making processes directly use this fact to produce near-optimal goodness-of-fit statistics and pixelized sky maps with high computational efficiency.

By applying this pipeline to data from the first observing season, we have demonstrated a key goal of the Pathfinder: that the noise level integrates down with time as expected for uncorrelated white noise. A careful analysis of the data and observing efficiencies obtained in Season 1 has allowed us to forecast the performance of the Pathfinder taking into account expected and already-implemented improvements to the instrument, analysis and observing strategy. Based on this forecast and on models for the CO emission at $z\sim3$, the Pathfinder expected to achieve a detection of the CO(1-0) auto power spectrum by the end of the 5-year observing campaign \citep{es_V}.

\begin{acknowledgments}
We thank graduated master student Erik Levén for his contribution to this work.
This material is based upon work supported by the National Science Foundation under Grant Nos.\ 1517108, 1517288, 1517598, 1518282 and 1910999, and by the Keck Institute for Space Studies under ``The First Billion Years: A Technical Development Program for Spectral Line Observations''.

Parts of the work were carried out at the Jet Propulsion Laboratory, California Institute of Technology, under a contract with the National Aeronautics and Space Administration, and funded through the internal Research and Technology Development program. DTC is supported by a CITA/Dunlap Institute postdoctoral fellowship. The Dunlap Institute is funded through an endowment established by the David Dunlap family and the University of Toronto. CD acknowledges support from an STFC Consolidated Grant (ST/P000649/1). 

JB, HKE, MKF, HTI, JGSL, MR, NOS, DW, and IKW acknowledge support from the Research Council of Norway through grants 251328 and 274990, and from the European Research Council (ERC) under the Horizon 2020 Research and Innovation Program (Grant agreement No.\ 819478, \textsc{Cosmoglobe}). 
JG acknowledges support from the University of Miami and is grateful to Hugh Medrano for assistance with cryostat design. 
SH acknowledges support from an STFC Consolidated Grant (ST/P000649/1). 
J.\ Kim is supported by a Robert A.\ Millikan Fellowship from Caltech. 
At JPL, we are grateful to Mary Soria for for assembly work on the amplifier modules and to Jose Velasco, Ezra Long and Jim Bowen for the use of their amplifier test facilities. 
HP acknowledges support from the Swiss National Science Foundation through Ambizione Grant PZ00P2{\_}179934. PCB is supported by the James Arthur Postdoctoral Fellowship. We thank Isu Ravi for her contributions to the warm electronics and antenna drive characterization. The Scientific color maps \texttt{roma} and \texttt{tokyo}~\citep{Crameri2018} are used in this study to prevent visual distortion of the data and exclusion of readers with color-vision deficiencies~\citep{Crameri2020}.
\end{acknowledgments}
\software{Matplotlib~\citep{matplotlib}; Astropy, a community-developed core Python package for astronomy~\citep{astropy}.}

\appendix
\begin{figure}
    \centering
    \includegraphics[width=\linewidth]{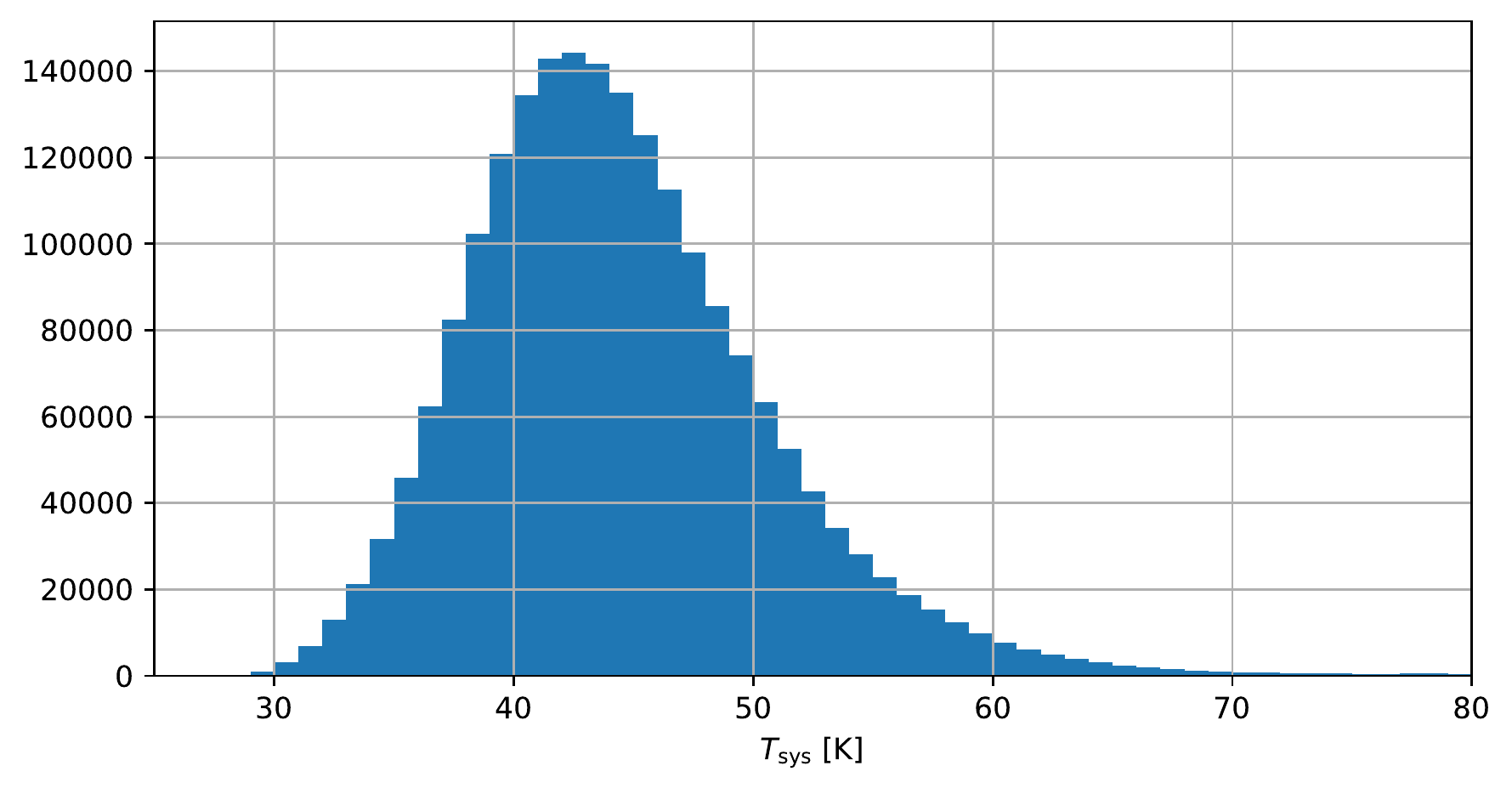}
    \caption{Histogram of sideband averaged system temperature, $T_\mathrm{sys}$, for all Season 1 observations.}
    \label{fig:tsys_range}
\end{figure}
\section{System temperature distribution}
Figure~\ref{fig:tsys_range} shows a histogram of the sideband averaged system temperature, $T_\mathrm{sys}$, for all Season 1 observations. Note that the sideband averages were taken after the frequency masking that happens during the low-level filtering. We also exclude observations at elevatons higher than 65 and lower than 35 degrees, as these are not used in the final analysis. We see that 95\,\% of the recorded system temperatures are in the range 34--60\,K, with a median value of 44\,K.

\bibliography{bibfile,early_science}{}
\bibliographystyle{aasjournal}

\end{document}